\documentclass[12pt]{article}
\usepackage{sty}
\pdfoutput=1

\begin{document}

\noindent {\bf \papertitle}

\noindent Ada~W.~C.~Yan$^1$ \\
\noindent Jie~Zhou$^2$ \\
\noindent Catherine~A.~A.~Beauchemin$^{3,4}$ \\
\noindent Colin~A.~Russell$^5$ \\
\noindent Wendy~S.~Barclay$^2$ \\
\noindent Steven Riley$^1$*

\begin{enumerate}
\item MRC Centre for Global Infectious Disease Analysis, Department of Infectious Disease Epidemiology, School of Public Health, Imperial College London, UK
\item Section of Virology, Department of Medicine, Imperial College London, UK
\item Department of Physics, Ryerson University, Canada
\item Interdisciplinary Theoretical and Mathematical Sciences (iTHEMS), RIKEN, Japan
\item Laboratory of Applied Evolutionary Biology, Department of Medical Microbiology, Academic Medical Center, University of Amsterdam, The Netherlands
\vspace{5mm}
\end{enumerate}
$^*$Corresponding author. Email address s.riley@imperial.ac.uk \\


\section*{Abstract}
When analysing \textit{in vitro} data, growth kinetics of influenza strains are often compared by computing their growth rates, which are sometimes used as proxies for fitness.  However, analogous to mechanistic epidemic models, the growth rate can be defined as a function of two parameters: the basic reproduction number (the average number of cells each infected cell infects) and the mean generation time (the average length of a replication cycle).  Using a mechanistic model, previously published data from experiments in human lung cells, and newly generated data, we compared estimates of all three parameters for six influenza A strains. Using previously published data, we found that the two human-adapted strains (pre-2009 seasonal H1N1, and pandemic H1N1) had a lower basic reproduction number, shorter mean generation time and slower growth rate than the two avian-adapted strains (H5N1 and H7N9).  These same differences were then observed in data from new experiments where two strains were engineered to have different internal proteins (pandemic H1N1 and H5N1), but the same surface proteins (PR8), confirming our initial findings and implying that differences between strains were driven by internal genes. Also, the model predicted that the human-adapted strains underwent more replication cycles than the avian-adapted strains by the time of peak viral load,  potentially accumulating mutations more quickly. These results suggest that the \textit{in vitro} reproduction number, generation time and growth rate differ between human-adapted and avian-adapted influenza strains, and thus could be used to assess host adaptation of internal proteins to inform pandemic risk assessment. 

\clearpage


\section*{Introduction}
Transmission experiments conducted using mammalian models are used to assess the pandemic risk of influenza virus strains currently circulating in animal reservoirs~\citep{Cox2014}. However, the cost and animal housing requirements for these experiments leads to small sample sizes~\citep{Nishiura2013} and scaling problems for systematic assessment across a large panel of strains. Animal experiment protocols are difficult to standardise, further complicating cross-study comparisons~\citep{Belser2018a}. On the other hand, \textit{in vitro} assays can be used to measure several factors affecting human adaptation of influenza strains, such as haemagglutinin receptor binding specificity, haemagglutinin pH of activation, and polymerase complex efficiency~\citep{Lipsitch2016}. However, methods to synthesise information from these assays and estimate strain-specific pandemic risk are lacking. If we could identify predictors for human adaptation which are quantitatively comparable between strains and can be measured using high-thoroughput assays, then their routine estimation would be highly informative for public health decision making.  In this study, we propose three such predictors: the basic reproduction number, the mean generation time, and the initial growth rate.  These parameters have been widely used to characterise the dynamics of epidemics, but only the last of these is commonly used in experimental virology to characterise \textit{in vitro}/within-host growth kinetics.

The initial growth rate is often directly interpreted in studies of virus evolution as a measure of \textit{in vitro} or within-host fitness~\citep{Sanjuan2010a,Lyons2018}.  In epidemiological studies, the initial growth rate can be expressed as a function of two parameters: the basic reproduction number and the mean generation time.  Each of these concepts can also be applied to the \textit{in vitro} or within-host context.

The basic reproduction number in an epidemiological context is the mean number of secondary infections due to an initial infected individual in an otherwise susceptible population. The cellular-level equivalent is the mean number of secondary infected cells due to an initial infected cell in an otherwise susceptible cell population.  Studies have estimated the basic cellular reproduction number for different pathogens, such as HIV/SHIV~\citep{Iwami2012,Iwami2015,Iwanami2017}, influenza A virus~\citep{Baccam2006} and rotavirus~\citep{Gonzalez-Parra2018}. A limited number of these studies have directly compared the basic reproduction number between different influenza or SHIV strains, as a measure of relative fitness~\citep{Mitchell2011,Iwanami2017,Farrukee2018}.

In an epidemiological context, the generation time is the time between infection of an individual and infection of a secondary case. The cellular-level equivalent is the time between infection of a cell and infection of a secondary cell.  Since a single cell will infect many secondary cells, this time varies between pairs of primary and secondary infected cells. A useful summary statistic for the distribution of these times is the mean generation time averaged over all secondary cells, in an otherwise susceptible cell population.    The mean generation time is an important parameter in models of pathogen evolution~\citep{Russell2012,Fonville2015,Illingworth2015,Nene2018,Geoghegan2016,Reperant2015}.  In the context of HIV, a mean generation time on the order of days has been linked to the rapid evolution of drug resistance, necessitating combination antiretroviral therapy~\citep{Perelson1996}.  Although mechanistic models have been used to quantify the mean generation time for HIV ~\citep{Perelson1996,Iwami2015,Althaus2009,Dixit2004}, the effects of between-strain differences on the dynamics of infection and evolution are not well studied.



This study aims to highlight the utility of the basic reproduction number and mean generation time in addition to the initial growth rate in considering the dynamics of an acute infection. We show that these three quantities (which we refer to collectively as cellular infection parameters) differ significantly between influenza virus strains. These differences correlate with different degrees of human adaptation, suggesting that these cellular infection parameters may be useful metrics to summarise biological differences. Importantly, although the parameters correlate with the \textit{in vivo} characteristic of human adaptation, they can be estimated using simple \textit{in vitro} assays.  We then use simulations to show the impact of these parameters on the number of replication rounds leading up to peak shedding; faster initial growth rates lead to a greater proportion of virions that were the product of several replication rounds.
We anticipate that strain differences in the number of replication cycles changes the accumulation rate of multiple mutations within a single infection, with consequences for the pace of virus evolution.

\clearpage
\section*{Results}

\begin{figure}[h]
\centering
\includegraphics[width = .8\textwidth]{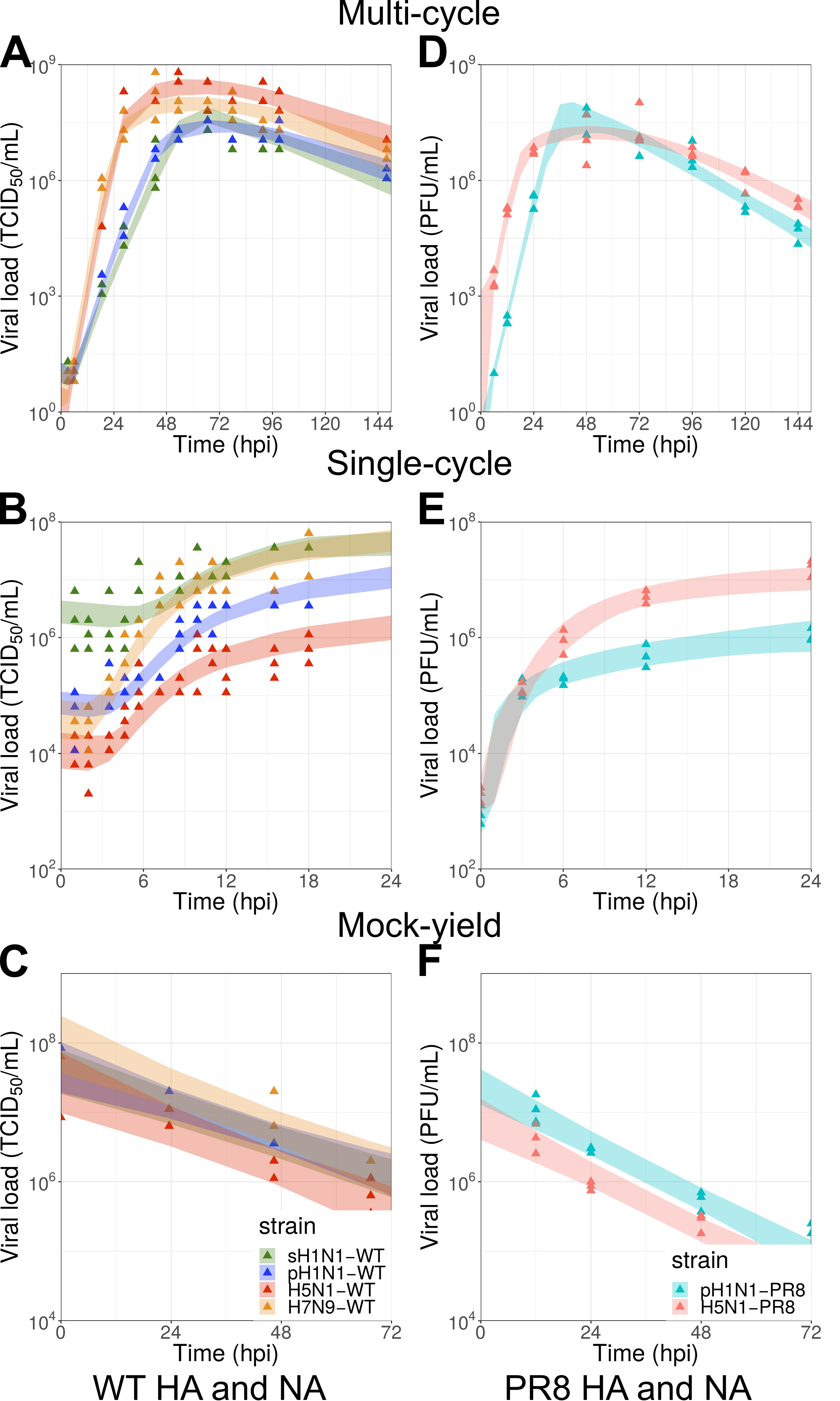}
\caption{
\textbf{Model fits to experimental data.}
The infectious viral load for the WT strains (A-C) and the strains with PR8 HA and NA (D-F). Fitted 95\% credible intervals are shown as shaded areas on top the the data (triangles). The infectious viral load is shown for (from top) the multi-cycle experiments, the single-cycle experiments and the mock-yield experiments.
}
\label{fig:data}
\end{figure}

We used a mechanistic model to characterise the \textit{in vitro} growth of four wild-type (WT) influenza A strains. These strains were \sHoneNone{} (a human-adapted strain, henceforth referred to as sH1N1-WT), \pHoneNone{} (human-adapted, pH1N1-WT), \HfiveNone{} (avian-adapted, H5N1-WT) and \HsevenNnine{} (avian-adapted, H7N9-WT).  The model contained the following processes: infection of a cell; an infected cell entering a state of virion production; production of infectious and non-infectious virions; infected cell death; loss of infectivity of free virions; and degradation of free infectious and non-infectious virions (see \methods{}).  We estimated the rates of these processes for the four strains by fitting the model to previously published data from single-cycle, multi-cycle and mock-yield assays~\citep{Simon2016}. The model accurately captured key features of the viral load (Fig.~\ref{fig:data}). With a low inoculum (MOI = 0.01), multiple generations of infection depleted the pool of susceptible cells over a period of 6 days resulting in an infectious viral load curve showing exponential growth, followed by a peak and exponential decay (Fig.~\ref{fig:data} A and D). With a larger inoculum (MOI = 3), most cells were infected immediately and the concentration of infectious virus plateaued earlier (Fig.~\ref{fig:data} B and E). The loss of infectivity of free virus was captured by the mock-yield assays where no cells were present (Fig.~\ref{fig:data} C and F). The model was also able to reproduce the observed total viral load patterns as quantified by qRT-PCR, although the viral load towards the end of the single-cycle experiments was slightly underestimated for the sH1N1-WT and pH1N1-WT strains (Fig. S1). 

\clearpage

\begin{figure}[t]
\includegraphics[width = \textwidth]{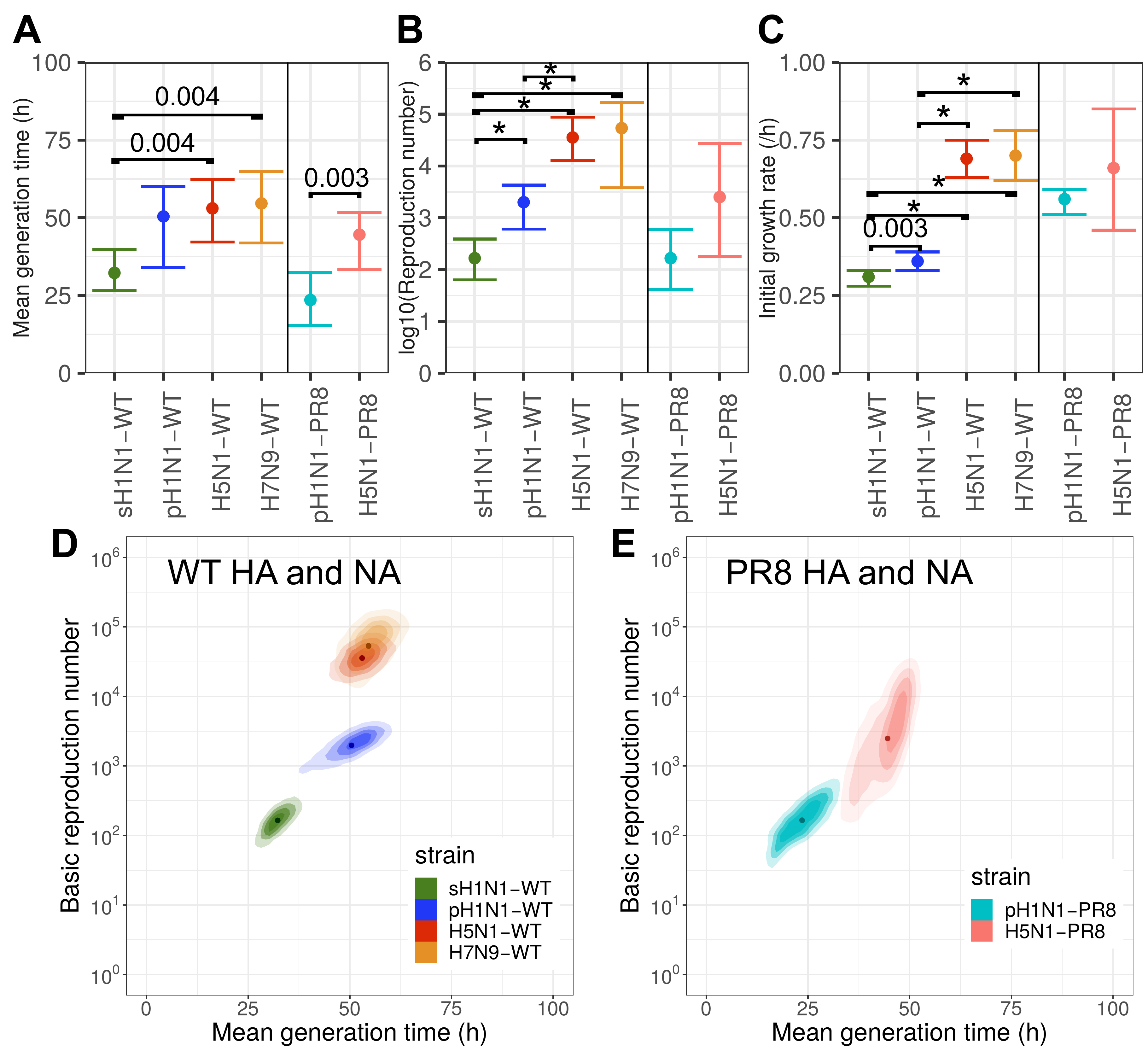}
\caption{
\textbf{Parameter estimates.}
(Top) The median and 95\% credible intervals for (A) the mean generation time, (B) basic reproduction number, and (C) initial growth rate, for the wild-type (WT) strains (left of each panel), and the strains with PR8 HA and NA (right).
Calculation of $p$-values is described in the \methods{}.
Statistically significant pairs are labelled ($\alpha = 0.05$ with Bonferroni correction for seven pairwise tests per parameter).
Asterisks denote $p < 0.001$.
(Bottom) Contour plots of the posterior density of the mean generation time and the basic reproduction number for each strain, for (D) the WT strains, and (E) the strains with PR8 HA and NA.
Fainter shading indicates lower support; dots indicate median values.
}
\label{fig:summary_stats}
\end{figure}

\clearpage
From the estimated rates of infection processes, we then calculated the basic reproduction number, the mean generation time and the initial growth rate (the `cellular infection parameters') for the four strains. We found that these parameters differed between human-adapted and avian-adapted strains, reflecting systematic differences in growth kinetics. We estimated the mean generation time to be between 25 and 65 hours for the four WT strains, ranging from 32 hours (median; 95\% CI 27~h--40~h) for sH1N1-WT to 55 h (median; 95\% CI 42~h--65~h) for H7N9-WT (Fig.~\ref{fig:summary_stats}A, left side of panel). There appeared to be an increasing trend in mean generation time from the strain most adapted to humans (sH1N1-WT) to the strain least adapted to humans (H7N9-WT). Strain differences were also observed in the basic reproduction numbers and initial growth rates (Figs.~\ref{fig:summary_stats}B-C). We note that combining data from three sets of experimental conditions (multi-cycle, single-cycle and mock-yield) enabled us to estimate these parameters accurately.
Fig.~S2 shows that parameter estimates using multi-cycle data only are less precise.
Data~S1 shows estimated values for the rates of all the infection processes in the model, and the correlations between them.

Fig.~\ref{fig:summary_stats}D further illustrates the observed parameter differences by showing that human-adapted and avian-adapted strains can be separated out when the estimated values of the basic reproduction number and the mean generation time are shown on a two-dimensional plot.  Avian-adapted strains (H5N1-WT and H7N9-WT) are in the top right corner of this plot; the most human-adapted strain, sH1N1-WT, is in the bottom left corner; and pH1N1-WT, which was an early pandemic strain and possibly not completely human-adapted, is in the centre.  We note the general trend that strains with a high basic reproduction number also tend to have a long mean generation time.  The same initial growth rate can be achieved via two different routes: a high basic reproduction number and a long mean generation time, or a low basic reproduction number and a short mean generation time~\citep{Nishiura2010,Wallinga2007}.  High basic reproduction numbers and short mean generation times would lead to very high initial growth rates, while low basic reproduction numbers and long mean generation times would lead to very slow initial growth rates; these extreme values may not be biologically plausible.  Because the basic reproduction number and the mean generation time are not completely independent, some types of data do not enable independent estimation of both parameters~\citep{Nishiura2010}.  This is not the case for our data.  For each strain, there is indeed some correlation in the posterior distribution between the basic reproduction number and the mean generation time (Fig.~S3; see also the diagonally stretched posterior distributions in Fig.~\ref{fig:summary_stats}D).  Despite this correlation, there is still sufficient information in the data to separate out estimates of both parameters for different strains (Figs.~\ref{fig:summary_stats}A-B, D).

We then investigated whether one of these three parameters alone (the basic reproduction number, mean generation time and initial growth rate) could capture the observed differences between strains.
We found that each parameter summarises changes in a different set of infection processes, and thus offers a complementary perspective on viral dynamics.  Fig.~S4 shows the sensitivity of each parameter to changes in the rates of underlying infection processes, as defined in the Materials and Methods.  For example, the leftmost bar shows the percentage change in the mean generation time for sH1N1-WT, if the infectivity of virions $\beta_{inf}$ were changed to that of H7N9-WT.  We see that the basic reproduction number is mostly affected by changes in the mean infectious period $\tau_I$ and the infectivity $\beta_{inf}$; the mean generation time is mostly affected by changes in $\tau_I$; and the initial growth rate is mostly affected by changes in $\beta_{inf}$.

We hypothesised that the differences between strains were driven by the internal proteins. To test this hypothesis, we used reverse genetics to engineer reassortant viruses that combined the haemagglutinin (HA) and neuraminidase (NA) segments of \PReight{} (PR8) with the remaining six gene segments from either a human adapted H1N1 virus (\ENGoneninefive{}) or an avian H5N1 virus (\Turkeyohfive{})~\citep{Li2018}; we refer to these reassortant strains as pH1N1-PR8 and H5N1-PR8 respectively. We conducted single-cycle, multi-cycle and mock-yield assays for these strains, and fitted the same model to the data to see if similar strain differences would be observed (Fig.~\ref{fig:data}D-F). The mean generation time for H5N1-PR8 was longer than that for pH1N1-PR8 (Fig.~\ref{fig:summary_stats}A, right side of panel). The median estimate of the basic reproduction number was also higher for H5N1-PR8 than pH1N1-PR8, although the difference was not statistically significant (Fig.~\ref{fig:summary_stats}B). Nevertheless, a positive correlation between basic reproduction number and mean generation time was observed (Fig.~\ref{fig:summary_stats}E). The initial growth rate did not appear to be different between the strains (Fig.~\ref{fig:summary_stats}C), but the estimate for H5N1-PR8 was imprecise because few observations were made during the exponential growth phase of the multi-cycle experiment, as the viral load plateaued earlier than expected (Fig.~\ref{fig:data}F).  Overall, the correlation between human adaptation and lower cellular infection parameter values (a lower basic reproduction number, shorter mean generation time and lower initial growth rate) was consistent with our previous results.  Note that we did not compare the cellular infection parameters directly between pH1N1-WT and pH1N1-PR8, or between H5N1-WT and H5N1-PR8, because our interest is in the difference between the internal proteins of pH1N1 and H5N1, rather than the changes introduced to each strain by changing their surface proteins.

\begin{figure}[t]
\centering
\includegraphics[width = \textwidth]{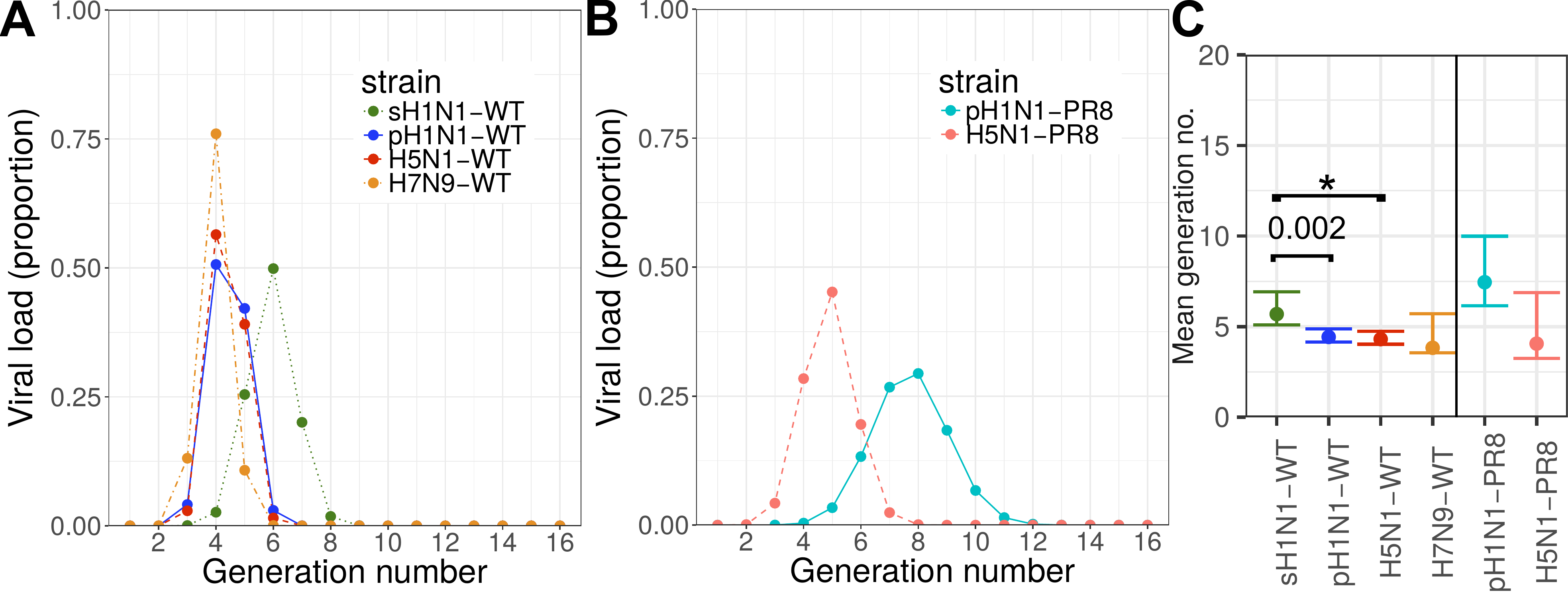}%
\caption{
\textbf{Simulated distribution of the number of virions in each generation, for the six experimental strains.} (A--B) The proportion of virions in each generation was calculated at the time of peak infectious viral load according to the maximum likelihood parameters for each strain, for (A) the WT strains, and (B) the strains with PR8 HA and NA.
As visual aids, lines join the number of virions at each discrete generation.
(C) The mean generation number was calculated for each sample from the joint posterior distribution for each strain.
The median and 95\% credible interval for the mean generation number are shown for the WT strains (left of each panel), and the strains with PR8 HA and NA (right).
Statistically significant pairs are labelled ($\alpha = 0.05$ with Bonferroni correction for seven pairwise tests per parameter).
Asterisks denote $p < 0.001$.}
\label{fig:gen_distribution}
\end{figure}

Simulations using the estimated parameters showed that by the peak time of infection, human-adapted strains compared to avian-adapted strains had a higher proportion of virions with a large generation number, which we define as one plus the number of replication cycles between the inoculum and the virion's production.  The time of peak infectious viral load is of interest because in an \textit{in vivo} infection, transmission is most likely around the time of peak viral load~\citep{Carrat2008}. To perform these simulations, we extended our model to associate each infectious virion and infected cell with a generation number (see \methods{}).  Fig.~\ref{fig:gen_distribution} shows the proportion of virions in each generation at the time of peak infectious viral load, according to the maximum likelihood parameter set for each strain, under multi-cycle experiment conditions. For these values, among the WT strains, sH1N1-WT had a higher proportion of virions with a high generation number compared to pH1N1-WT, H5N1-WT and H7N9-WT (Fig.~\ref{fig:gen_distribution}A). pH1N1-PR8 also had a higher proportion of virions with a high generation number compared to H5N1-PR8 (Fig.~\ref{fig:gen_distribution}B). To summarise this distribution, we calculated the mean generation number at the time of peak infectious viral load.  Calculating this statistic across the joint posterior distribution confirmed that human-adapted strains had a higher mean generation number compared to avian-adapted strains, and thus virions were on average a product of more replication cycles (Fig.~\ref{fig:gen_distribution}C).  

\begin{figure}[t]
\centering
\includegraphics[width = \textwidth]{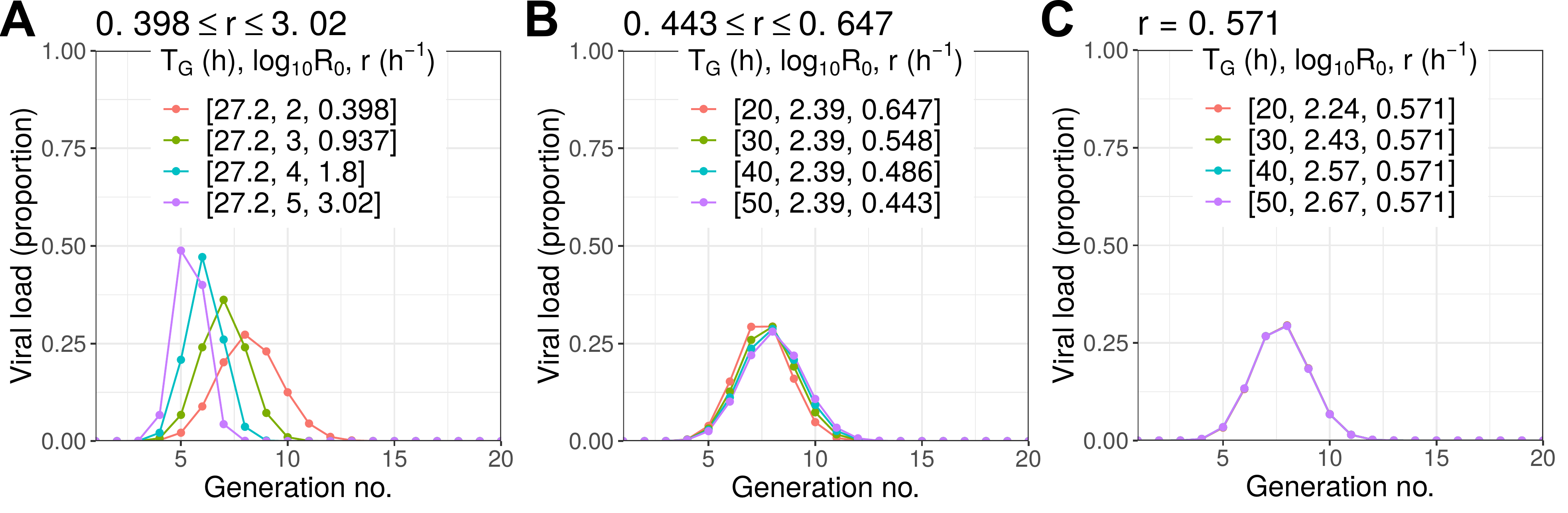}%
\caption{
\textbf{Simulated distribution of the number of virions in each generation, as cellular infection parameters are changed systematically.} The proportion of virions in each generation (the generation number distribution) was calculated at the time of peak infection.
The cellular infection parameters in the legend were changed to the values shown.  In each panel, one parameter is held constant: the mean generation time $T_G$ (A), basic reproduction number $R_0$ (B), or initial growth rate $r$ (C).
}
\label{fig:gen_distribution_systematic}
\end{figure}

We investigated the hypothesis that longer generation times alone might be driving the difference in number of replication cycles at peak viral shedding. However, we found that this was not the case. A sensitivity analysis showed that out of the three parameters, changing the initial growth rate had the largest effect on the generation number distribution at the time of peak infectious viral load (Fig.~\ref{fig:gen_distribution_systematic}). When we held the mean generation time constant (Fig.~\ref{fig:gen_distribution_systematic}A), we were able to achieve very different generation number distributions by varying the other parameters. However, when we held the initial growth rate constant (Fig.~\ref{fig:gen_distribution_systematic}C), varying the other parameters did not substantially change the generation number distribution. Hence, the initial growth rate was the main driver of the generation number distribution.

We conducted a number of sensitivity analyses. We tested different model structures to see the impacts of model assumptions on estimated parameter values. First, we ignored loss of free virions due to entry into target cells (\simethodsfilename{}). These changes only changed estimated parameter values slightly, and did not affect the relationship between the cellular infection parameters, human adaptation, and the generation number distribution (Fig.~S5). Second, we decreased the level of heterogeneity in the generation time.  We first decreased heterogeneity by narrowing the distribution of the latent period and duration of virion production, by setting $n_L = n_I = 60$ in the model equations.  This change also did not affect the above relationships (Fig.~S6).  

We then decreased heterogeneity by changing the timing of production of virions.  In the model in the main text, once each cell starts producing virus, it does so at a constant rate until death.  (Note that due to heterogeneity in infected cells' lifetimes, the bulk production rate of virions by many cells varies over time.) On the other hand, models including intracellular processes predict an age-dependent production rate~\citep{Heldt2013}.  Fig.~S7 shows the viral load estimated using a model where the production rate of virions increases over an infectious cell's age of infection  (\simethodsfilename{}). In this model, most virions are produced toward the end of an infected cell's lifespan, so there is less heterogeneity in the generation time.  This model predicts slightly different dynamics from the model in the main text (Fig.~\ref{fig:data}).
For the multi-cycle data, the new model predicts a sharp increase in viral load followed by a sharp decline for all strains.
The model in the main text predicted similar dynamics for human-adapted strains, but a gradual plateau in viral load followed by a decrease for avian-adapted strains.
This difference can be seen by inspection of the data, and is missed by the new model.
For the single-cycle data, the new model produces "bumpy" curves with multiple stages of rapid viral growth, whereas the model in the main text produces the sigmoidal curves characteristic of single-cycle growth kinetics.
Model predictions of the mock-yield viral load have a different slope from the data, particularly for H5N1-PR8.
We hypothesise that in the new model, the slope of the multi-cycle viral load during its decay phase is less driven by the infected cell lifespan and more driven by the loss of virus infectivity; hence, the new model is unable to explain strain differences in the multi-cycle viral load decay rate without introducing artificial differences in the rate of loss of virus infectivity.
Due to these qualitative differences between the data and the estimated viral load, we conclude that the model in the main text, where the production rate is constant, is more appropriate.

Last, we removed heterogeneity in the generation time altogether, which we hypothesised would change the relationship between the cellular infection parameters and the number of replication cycles by the time of peak infectious viral load. We conducted this sensitivity analysis because evolutionary models often assume a fixed generation time, such that at a given time all virions belong to the same generation~\citep{Russell2012,Geoghegan2016,Illingworth2015}.  Fig.~S8 shows the simulated generation number of virions at the time of peak infectious viral load for a model with a fixed generation time; details are given in \simethodsfilename{}. When we held the basic reproduction number constant (Fig.~S8B), varying the other parameters did not change the generation number, whereas when we held the generation time or the initial growth rate constant, we were able to change the generation number by varying the other parameters. Hence, for the fixed generation time model, only the basic reproduction number influences the generation number at the time of peak viral load.  This drastically different result implies that quantitative predictions of the number of mutations accumulated during the time course of infection should account for heterogeneity in the generation time.  This heterogeneity arises because virions are produced throughout the infectious period of a cell, and the infected cell lifespan is itself highly heterogeneous~\citep{Holder2011,Beauchemin2017}.  These sources of heterogeneity are reflected in our model structure, and lead to successive generations of virions overlapping in time rather than all virions belonging to the same generation.  

\clearpage
\section*{Discussion}

In this study, we quantified \textit{in vitro} parameters --- the basic reproduction number, the mean generation time, and the initial growth rate --- for six influenza strains using a mechanistic model. Avian-adapted strains had higher initial growth rates compared to human-adapted strains, which can be directly seen in the data.  They were also estimated to have higher basic reproduction numbers and longer mean generation times. We found that these differences were driven by differences in internal proteins, suggesting a role for experiments to measure these parameters in assessing pandemic risk due to adaptation of internal proteins.

The main limitation of of our results with respect to this finding is that we have only assessed these parameters in six strains, three with human-adapted internal genes and three with avian-adapted internal genes. We note that experiments for one human-adapted strain (pH1N1-PR8) and one avian adapted strain (H5N1-PR8) were designed specifically to measure these parameters and test the hypotheses generated using data from the other strains. Nonetheless, to strengthen the case for the the utility of these parameters, it would be desirable to repeat these experiments for a wider panel of strains with the same surface proteins, to see whether differences in cellular infection parameters still hold across more human-adapted and avian-adapted strains. A positive result would strengthen the case for the routine use of these assays to assess pandemic risk. 

The strength of comparing strains engineered to have the same surface proteins is that we could observe differences due to internal proteins which may otherwise have been obscured by receptor binding differences.  Differences in internal proteins are likely to generalise across cell types and to conditions in epithelial cells lining the respiratory tract.
Differences in surface proteins, such as those which affect receptor binding, are more likely to be cell-type specific. Hence, if considering the combined effect of internal and surface proteins in wild-type strains, experimental conditions would need to more closely mimic conditions in the respiratory tract, for example by using primary differentiated human epithelial cells rather than the A549 cell line.  We caution that the observed differences in reproduction number, generation time and growth rate due to internal protein adaptation need to be considered together with surface protein adaptations for comprehensive pandemic risk assessment.
For example, a previous study used primary normal human bronchial epithelial cells as a model of the respiratory tract, and found that higher, not lower, basic reproduction numbers were associated with human adaptation~\citep{Mitchell2011}.  Although this result appears to contradict our findings, higher basic reproduction numbers for human-adapted strains in primary normal human bronchial epithelial cells were likely due to strain differences in receptor binding specificity, since these cells primarily express $\alpha$2,6-linked sialic acid receptors which are already well understood to be favoured by human-adapted strains. In our experiments where strains were engineered to have the same surface proteins, we could eliminate the already-known effects of receptor binding and isolate the effect of differences in internal proteins.

We suggest the growth rate, the basic reproductive number, and the generation time taken together may form a useful description of influenza strain growth dynamics in the absence of immunity. Uncovering the mechanisms leading to the observed differences may lead to increased understanding of the biology of host species jumps. Intuitively, one may expect that increasing the growth rate through both increasing the reproduction number and shortening the generation time would give the greatest evolutionary advantage; instead, human adaptation favours a shorter generation time but also a smaller cellular reproduction number. This result suggests a potential trade-off between reproduction number and generation time that may enable different replication strategies in the human respiratory tract and the avian gastrointestinal tract, the primary sites of infection for human and avian infection respectively.  The human respiratory tract may elicit a stronger local immune response, favouring strains which can reproduce quickly and release new virions which spread and evade the local response.  The conditions for free virion survival are different in the avian gastrointestinal tract, which could favour a higher reproduction number to maximise the contribution of each infectious cell to onwards replications.

As reviewed by Wargo \textit{et al.}~\citep{Wargo2012}, virological data analysis protocols have mostly used the initial growth rate to measure \textit{in vitro} fitness, interpreting a larger growth rate as a fitness advantage.  The previous analysis of four of the strains presented here by Simon \textit{et al.}~\citep{Simon2016} showed strain differences in the rate of infection, which is closely related to the initial growth rate. A limited number of studies have computed the basic reproduction number, and fewer still have compared it between virus strains~\citep{Mitchell2011,Iwanami2017,Farrukee2018}.   
Some modelling studies have computed another related quantity, the infecting time, which is the time between a cell starting to produce virus and infection of the \textit{first} secondary cell~\citep{Pinky2016,Paradis2015a,Holder2011a,Holder2011,Pinilla2012,Petrie2013,Petrie2015,Gonz2018}. The mean generation time differs from the infecting time because the mean generation time is averaged over \textit{all} secondary cells, and also includes the period before an infected cell starts producing virus.  If cell and virion loss are assumed to be negligible during early infection, then the initial growth rate is inversely proportional to the sum of the infecting time and the latent period before an infected cell produces virus.

In our study, we showed that a lower basic reproduction number, a shorter mean generation time and a slower initial growth rate are associated with human adaptation.  However, we also showed that strain differences in the three cellular infection parameters were driven by changes in rates of different underlying infection processes, such that each parameter provides a different perspective.  Some studies have directly compared the rates of these infection processes for strains differing by a single mutation~\citep{Pinilla2012,Holder2011a,Petrie2015,Simon2016}. Because mutations may affect one of more infection processes, combining these processes into a smaller number of summary measures --- the basic reproduction number, mean generation time, and initial growth rate --- may enable easier between-strain comparisons.

We also found that strains with lower initial growth rates have a higher proportion of virions which are a product of a large number of replication cycles at the time of peak infection, when transmission is most likely to occur for the \textit{in vivo} case. For a given mutation rate per replication cycle (as estimated by Parvin \textit{et al.}~\citep{Parvin1986} and Nobusawa \textit{et al.}~\citep{Nobusawa2006}), this implies quicker accumulation of mutations. The degree of heterogeneity in the generation time also changed the generation number distribution. If we unrealistically assumed no heterogeneity in the generation time, then the generation number at the time of peak viral load depended entirely on the basic reproduction number rather than the initial growth rate, highlighting the importance of heterogeneity in the generation time in evolutionary models.

A caveat of this second finding is that the relationship between the generation number distribution and the initial growth rate would be different \textit{in vivo} due to a time-dependent immune response.  However, the overall finding that different cellular infection parameter values lead to different generation number distributions at a given time should still hold.  Also, in linking the accumulation of generations to accumulation of mutations, we have assumed that the number of errors introduced between primary and secondary virions is independent of when the secondary virion was produced during the infected cell's lifespan.
However, the number of errors introduced may increase with the age of the infected cell, since a virion exiting an `older' cell may have been a product of more intracellular replication and transcription cycles than a virion exiting a `younger' cell.   In our model, we have not considered multiple intracellular replication and transcription cycles before virus release.  However, the relationship between a cell's age and the number of mutations in produced virions is yet to be well understood.

\section*{\methods{}}

\subsection*{Strains with wild-type HA and NA}

In brief, \sHoneNone{}, \pHoneNone{}, \HfiveNone{} and \HsevenNnine{} stocks were grown in MDCK cells.
Single-cycle and multi-cycle experiments were carried out in triplicate in A549 human lung carcinoma cells in T25 flasks.
Cells were infected at MOI = 3 and MOI = 0.01 for the single-cycle and multi-cycle experiments respectively, for an incubation period of 1 hour at 37$^\circ$C.
For the single-cycle experiment, the cells were washed with an acidic saline wash after the incubation period.
At set time points, 0.5 mL of the 10 mL supernatant volume was harvested and replaced with 0.5 mL of fresh media.
A mock-yield experiment was also performed in triplicate, where $10^7$ pfu of each strain was left to decay in cell-free media incubated at 37$^\circ$C.
Infectious virus was quantified using a TCID$_{50}$ assay, and for the single-cycle and multi-cycle experiments, total virus was quantified using qRT-PCR.
See Simon \textit{et al.}~\citep{Simon2016} for the full protocol.

\subsection*{Strains with PR8 HA and NA}

Human embryonic kidney (293T) (ATCC), human lung adenocarcinoma epithelial cells (A549) (ATCC) and Madin-Darby canine kidney (MDCK) cells (ATCC) cells were maintained in Dulbecco's modified Eagle's medium (DMEM; Gibco, Invitrogen) supplemented with 10\% fetal calf serum, 1\% non-essential amino acids and 1\% penicillin-streptomycin (5000 IU/mL; 5000 $\mu$L) at 37$^\circ$C and 5\% CO$_{2}$.

We generated two reverse genetics viruses (pH1N1-PR8 and H5N1-PR8). In these viruses, the HA and NA genes were from the laboratory adapted strain A/Puerto Rico/8/34 (H1N1), and the six remaining gene segments were from either A/England/195/2009 (pH1N1) or A/Turkey/05/2005 (H5N1). Eight poll plasmids encoding the indicated virus segments and four helper expression plasmids encoding polymerase components and NP expressed by the pCAGGS vector were transfected into 293T cells. 
After 24 hours, the transfected 293T cells were resuspended and co-cultured with MDCK cells.
Virus stocks were thus grown on MDCK cells using serum free DMEM supplemented with 1$\mu$g/mL of TPCK trypsin (Worthington). Viruses were stored in -80$^\circ$C and titrated on MDCK cells by plaque assay to determine the dilution required to achieve a given multiplicity of infection.

All infection experiments were performed in triplicate (three wells on the same plate).
For the single-cycle and multi-cycle experiments, A549 cells were plated in a six-well plate ($2.5 \times 10^6$ cells per well).
One day after plating, medium was removed from cells and cells were washed twice with PBS (3 mL/well), then covered with 500 $\mu$L serum-free DMEM medium.
Cells were infected at MOI = 5 and MOI = 0.01 for the single-cycle and multi-cycle experiments respectively; the virus was thawed in a 37 $^\circ$C water bath, then diluted with serum-free DMEM to a volume of 500 $\mu$L.
After an incubation period of one hour, the inoculum was removed, cells were washed four times with 3 mL serum-free DMEM, and 3 mL serum-free DMEM with 1 $\mu$g/mL TPCK-treated trypsin was added to each well (without cells). At each measurement time (shown in Fig.~\ref{fig:data}), the plates were shaken and 300 $\mu$L supernatant was collected, and replaced by 300 $\mu$L serum-free DMEM with 1 $\mu$g/mL TPCK-treated trypsin.
The supernatant was frozen at -80$^\circ$C for later quantification.
For the mock-yield experiments, 3 mL of virus at concentration $10^7$ pfu/mL was added to each well. At each measurement time, 300 $\mu$L supernatant was collected without replacement and frozen at -80$^\circ$C for later quantification.

Plaque assays were carried out in confluent monolayers of MDCK cells in 12-well plates.
100 $\mu$L of each tenfold virus dilution was applied to each cell and incubated for 1 hour at 37$^\circ$C.
The inoculum was then removed, and the cells were overlaid with 0.6\% agarose in MEM including 1 $\mu$g/mL TPCK-treated trypsin and 0.3\% bovine serum albumin fraction V (Gibco).
The cells were then incubated at 37$^\circ$C.
After 3 days, the agarose was removed and the cells stained with 1 mL 0.5\% crystal violet.

\subsection*{Mathematical model}

\begin{figure}[tbhp]%
\centering
\includegraphics[width = .4\textwidth]{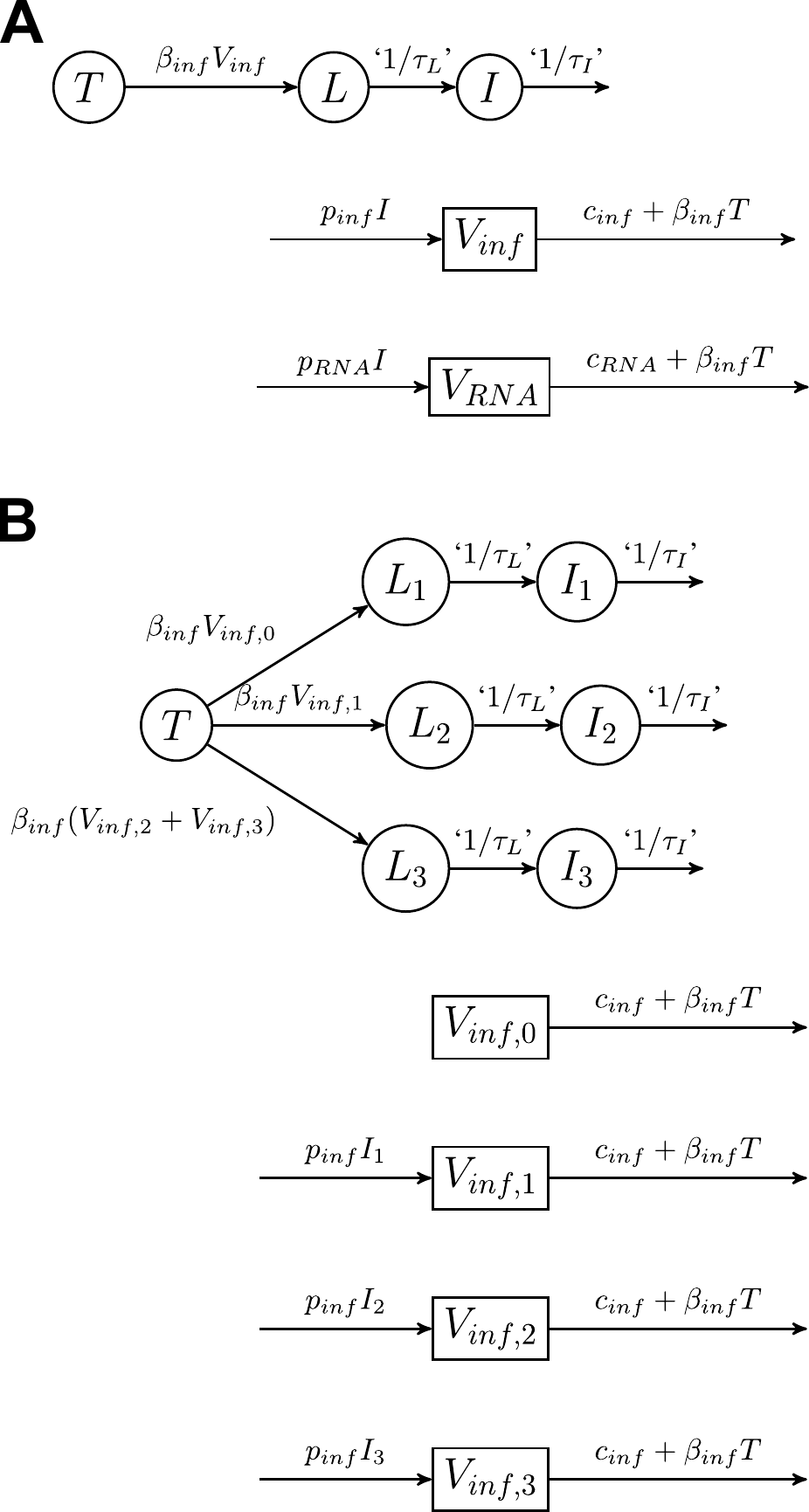}%
\caption{\textbf{(A) Viral dynamics model; (B) Model tracking generation number.} 
Circles indicate cell compartments, while squares denote virion compartments.
Quotation marks indicate that the waiting time between compartments is gamma-distributed rather than exponentially distributed, and that the mean waiting time is the inverse of the `rate' given.
(B) illustrates $N = 3$ generations.}
\label{fig:model}
\end{figure}

Our mechanistic model (Fig.~\ref{fig:model}) included the following processes: 

\begin{itemize}
\item infection of a cell at rate $\beta_{inf}$; 
\item an infected cell entering a state of virion production ($\tau_L$ is the mean time spent in the latent period between infection and the state of virion production); 
\item production of infectious and non-infectious virions at rates $p_{inf}$ and $p_{RNA}$ respectively; 
\item infected cell death ($\tau_I$ is the mean time from entering the state of virion production to death); 
\item loss of infectivity of free infectious virions at rate $c_{inf}$; and
\item degradation of free infectious and non-infectious virions at rate $c_{RNA}$.
\end{itemize}

This model was adapted from that used by Simon \textit{et al.}~\citep{Simon2016}.
We did not model time-dependent effects of the immune response.
The model is formulated as a set of ordinary differential equations, as shown in the \simethodsfilename{}.

Single-cycle, multi-cycle and mock-yield experiments can be simulated using this model by changing the initial conditions.
Single-cycle and multi-cycle experiments start with a fixed number of target cells $T_0$, and infectious and total virus at a low MOI (multi-cycle) or a high MOI (single-cycle).
Mock-yield experiments start with no target cells and a large amount of infectious virus.

The cellular infection parameters were expressed as functions of the rates of infection processes.
The basic reproduction number is
\begin{equation}
R_0 = \frac{\beta_{inf} T_0 p_{inf} \tau_I}{c + \beta_{inf} T_0};
\label{eq:R_0}
\end{equation}

the mean generation time, which we consider in the context of a fully susceptible population, is
\begin{equation}
\tau_G = \tau_L + \frac{n_I + 1}{2n_I}\tau_I + \frac{1}{c + \beta_{inf} T_0};
\label{eq:gen_time}
\end{equation}

and the initial growth rate $r$ is computed by linearising around the disease-free equilibrium $[T, L_i, I_i, V_{inf}] = [T_0, 0, 0, 0]$~\citep{Nowak1997,Lee2009a}.

To relate the data to the viral load predicted by the model, we introduced parameters to convert the number of infectious and total virions to the concentration of infectious and total virus measured in the supernatant.
We then modelled observation error as lognormal, with an observation threshold at low titres.

\subsection*{Parameter estimation}

Parameters were estimated for each strain separately using an adaptive Metropolis-Hastings algorithm.
For a given strain, parameters were estimated using combined data from the multi-cycle, single-cycle and mock yield experiments.
For a given strain, these experiments share all model parameters except for the production rate of infectious virus $p_{inf}$, and otherwise differ only in initial conditions.
$p_{inf}$ for the single-cycle experiments is modelled as lower (or equal to) that for the multi-cycle experiments, to account for decreased infectious viral production due to the presence of defective interfering particles at a high multiplicity of infection~\citep{Simon2016}.

The viral RNA degradation rate, shape parameters of the latent period and viral production period distributions, initial number of target cells, supernatant volume and observation threshold were fixed.
All other parameter values, including initial conditions, were estimated.
A multi-dimensional uniform distribution was used as the prior, with some parameters log-transformed.
Tables showing the fixed parameter values and the priors for the estimated parameters are included in the \simethodsfilename{}.

\subsection*{Statistical analysis}

For each pair of viruses in the study, for each estimated parameter, we computed the ratio of the virus-specific parameter values, and tested for deviations of the ratio from 1.
For each parameter, $p$-values for each virus pair were computed by sampling with replacement from each marginal posterior distribution, and letting $q$ be the proportion of sampled pairs whose ratio exceeds 1. 
The $p$-value was then $p = 2(min(q, 1 - q))$. 
The strains with wild-type HA and NA were only compared to each other, and not to the strains with PR8 HA and NA.

\subsection*{Model with generation numbers}

We extended our model to explicitly associate each virion and infected cell with a generation number.
The inoculum was denoted virus generation 1, the cells infected by the inoculum denoted cell generation 1, the infectious virions produced by those cells denoted virus generation 2, and so forth.
This model enables us to calculate the proportion of virions from each generation at a given time.
The model with generations is illustrated in Fig.~\ref{fig:model}B for tracking $N = 3$ generations; the actual implementation tracks 20 generations.
(Cells and virions above generation $N$ are lumped into the generation $N$ compartments).
The bulk dynamics of the model remain the same, that is, the number of cells in $L$ in Fig.~\ref{fig:model}A is equal to the sum of the numbers of cells in $L_1$, $L_2$ and $L_3$ in Fig.~\ref{fig:model}B, and similarly for $I$ and $V_{inf}$.  Model equations are in the \simethodsfilename{}.

Data and code to reproduce all results can be found at \repo{}.

\section*{Acknowledgments}

This work was supported by a Wellcome Trust Collaborator Award (UK, grant 200187/Z/15/Z, supporting AWCY, JZ, CAR, WSB and SR). This work was also supported in part by a Wellcome Trust Investigator Award (UK, grant 200861/Z/16/Z, supporting SR), a Discovery Grant 355837-2013 from the Natural Sciences and Engineering Research Council of Canada
(www.nserc-crsng.gc.ca), and by Early Researcher Award ER13-09-040 from
the Ministry of Research and Innovation and Science of the Government of
Ontario (www.ontario.ca/page/early-researcher-awards), both awarded to
CAAB, and by Interdisciplinary Theoretical and Mathematical Sciences
(iTHEMS, ithems.riken.jp) at RIKEN (CAAB).

\section*{References}

\bibliographystyle{apalike}
\bibliography{Mendeley,unpublished}

\end{document}


\noindent {\bf \papertitle: Supplementary Text}

\noindent Ada~W.~C.~Yan$^1$ \\
\noindent Jie~Zhou$^2$ \\
\noindent Catherine~A.~A.~Beauchemin$^{3,4}$ \\
\noindent Colin~A.~Russell$^5$ \\
\noindent Wendy~S.~Barclay$^2$ \\
\noindent Steven Riley$^1$*

\begin{enumerate}
\item MRC Centre for Global Infectious Disease Analysis, Department of Infectious Disease Epidemiology, School of Public Health, Imperial College London, UK
\item Section of Virology, Department of Medicine, Imperial College London, UK
\item Department of Physics, Ryerson University, Canada
\item Interdisciplinary Theoretical and Mathematical Sciences (iTHEMS), RIKEN, Japan
\item Laboratory of Applied Evolutionary Biology, Department of Medical Microbiology, Academic Medical Center, University of Amsterdam, The Netherlands
\vspace{5mm}
\end{enumerate}
$^*$Corresponding author. Email address s.riley@imperial.ac.uk \\

\clearpage


\section*{Details of mathematical model}\label{sec:model}

\subsection*{Model in units of virions}

The model equations are based off those in the study by Simon \textit{et al.}~\citep{Simon2016} with modifications to account for loss of virus due to entry into target cells~\citep{Handel2007}.
The model structure was first used by Perelson \textit{et al.}~\citep{Perelson1996} to model HIV infection, and first applied to influenza infection by Baccam \textit{et al.}~\citep{Baccam2006}.
Target cells (uninfected cells) ($T$) are infected by free virus ($V_{inf}$) according to a mass-action term with rate $\beta_{inf}$.
The newly infected cells ($L$) enter a latent period before they can produce free virus.
At the end of the latent period, the cells transition from the $L$ to the $I$ stage, and the infectious cells produce free virus at a constant rate $p_{inf}$.
Free virus decays at a constant rate $c$, and infectious cell lifespans are Erlang-distributed with mean $\tau_I$ and shape parameter $n_I$.

\begin{subequations}
\begin{align}
\frac{dT}{dt} &= -\beta_{inf} T V_{inf},\\
\frac{dL_{1}}{dt} &= \beta_{inf} T V_{inf} - \frac{n_{L}}{\tau_{L}} L_{1},\\
\frac{dL_{i}}{dt} &= \frac{n_{L}}{\tau_{L}} (L_{i-1} - L_{i}),\\
\frac{dI_{1}}{dt} &= \frac{n_{L}}{\tau_{L}} L_{i} - \frac{n_{I}}{\tau_{I}} I_{1},\\
\frac{dI_{i}}{dt} &= \frac{n_{I}}{\tau_{I}} (I_{i-1} - I_{i}),\\
\frac{dV_{inf}}{dt} &= p_{inf} \sum_{i=1}^{n_{I}} I_{i} - c V_{inf} - \beta_{inf} T V_{inf}
\end{align}
\label{eq:model}
\end{subequations}

\begin{tabular}{rrrr}
  \hline
Parameter & Definition & Units \\ 
  \hline
  $c$ & Free virus decay rate & h$^{-1}$\\ 
  $\beta_{inf}$ & Rate at which cells are infected by virus & infectious unit $^{-1}$ h$^{-1}$ \\
  $\tau_L$ & Mean latent period & h \\ 
  $\tau_I$ & Mean infectious period & h \\ 
  $p_{inf}$ & Rate at which virus is produced by infected cells & infectious unit (cell$^{-1}$) h$^{-1}$ \\ 
  $n_L$ & Number of latent stages & -\\
  $n_I$ & Number of infectious stages & -\\
   \hline
\end{tabular}

The model can be written as a set of ordinary differential equations (Eq.~\ref{eq:model}). 
In Eq.~\ref{eq:model}, $V_{inf}$ is in infectious units, which we define as the amount of free virus lost from the medium to infect a single cell.
We will shortly derive equations to link $V_{inf}$ to experimentally measured units.
The infected cells pass through $n_L$ latent stages and $n_I$ infectious stages, where the duration of each stage is exponentially distributed.
As a result, the latent and infectious periods are Erlang distributed.
Holder \textit{et al.}~\citep{Holder2011} showed that for influenza, compared to exponentially distributed latent and infectious periods, normally distributed or lognormally distributed latent and infectious periods lead to better fits to single-cycle data.
When the shape parameter is much greater than 1, the Erlang distribution is similar in shape to the normal and lognormal distributions.
The Erlang distribution was also used in influenza viral dynamics models~\citep{Pinilla2012,Beggs2015,Petrie2013,Liao2016,Paradis2015a}, and in HIV viral dynamics models~\citep{Beauchemin2017}.

\subsection*{Linking model to experimentally observed quantities}

To link our model to observed experimental quantities, we make two modifications.
In Eq.~\ref{eq:model}, $V_{inf}$ is in infectious units, which we define as the amount of free virus lost from the medium to infect a single cell.
However, for the strains with PR8 HA and NA, we observe the number of plaque forming units (pfu) per mL of supernatant.
We define $n$ to be the number of infectious units per pfu, where $n \geq 1$, and $S$ to be the supernatant volume.  
Then we can make the following unit conversions:

\begin{subequations}
\begin{align}
V_{pfu} &= \frac{V_{inf}}{n},\\
V_{pfu/mL} &= \frac{V_{inf}}{nS},\\
\beta_{pfu/mL} &= \beta_{inf}nS, \\
p_{pfu/mL} &= \frac{p_{inf}}{nS},
\end{align}
\label{eq:conversion}
\end{subequations}
to arrive at the model equations in observation units:

\begin{subequations}
\begin{align}
\frac{dT}{dt} &= -\beta_{pfu/mL} T V_{pfu/mL},\\
\frac{dL_{1}}{dt} &= \beta_{pfu/mL} T V_{pfu/mL} - \frac{n_{L}}{\tau_{L}} L_{1},\\
\frac{dL_{i}}{dt} &= \frac{n_{L}}{\tau_{L}} (L_{i-1} - L_{i}),\\
\frac{dI_{1}}{dt} &= \frac{n_{L}}{\tau_{L}} L_{i} - \frac{n_{I}}{\tau_{I}} I_{1},\\
\frac{dI_{i}}{dt} &= \frac{n_{I}}{\tau_{I}} (I_{i-1} - I_{i}),\\
\frac{dV_{pfu/mL}}{dt} &= p_{pfu/mL} \sum_{i=1}^{n_{I}} I_{i} - c V_{pfu/mL} - \beta_{inf} T V_{pfu/mL}.
\end{align}
\label{eq:model_obs}
\end{subequations}

\begin{tabular}{rrrr}
  \hline
Parameter & Definition & Units \\ 
  \hline
  $\beta_{pfu/mL}$ & Rate at which cells are infected by virus & (pfu/mL)$^{-1}$ h$^{-1}$ \\ 
  $p_{pfu/mL}$ & Rate at which virus is produced by infected cells & (pfu/mL) (cell$^{-1}$) h$^{-1}$ \\ 
  $n$ & Number of infectious units per pfu & infectious unit (pfu)$^{-1}$ \\
  $S$ & Supernatant volume & mL\\ 
   \hline
\end{tabular}

For the strains with wild-type HA and NA, we observe the number of TCID$_{50}$s per mL of supernatant.
We use the same equations where $n$ is now the number of TCID$_{50}$ per infectious unit, where $n \geq 1$.

For the strains with wild-type HA and NA, the total free viral load (in RNA copy number/mL) was measured in addition to the infectious viral load.
The total viral load differs from the infectious viral load because an infected cell produces both infectious and non-infectious virions, the latter of which arise due to defects introduced during the viral replication process~\citep{Nayak1985,Marriott2010}.  
It had previously been shown that fitting a model to the infectious and total viral load improves parameter estimates compared to fitting to infectious viral load only~\citep{Petrie2013}, and several other studies have fitted models including the total viral load to data~\citep{Pinilla2012,Paradis2015a,Beggs2015}.
We model production of total virus ($V_{RNA}$) by infectious cells ($I$) at a constant rate $p_{RNA}$.
Total virus decays at a constant rate $c_{RNA}$, and is also lost due to entry into target cells.  
We assume that one infectious unit contributes one copy number to the total viral load~\citep{Yan2019}.
The equation for the total viral load concentration ($V_{RNA/mL}$) is then
\begin{equation}
\frac{dV_{RNA/mL}}{dt} = p_{RNA/mL} \sum_{i=1}^{n_{I}} I_{i} - c_{RNA} V_{RNA/mL} - \beta_{inf}n T V_{pfu/mL}.
\label{eq:model_rna}
\end{equation}

For the multi-cycle experiments with wild-type HA and NA, infection begins with the addition of virus to target cells and medium at $t = 0$.
The initial conditions are then $T(0) = T_0$, $V_{MC, TCID50/mL}(0) = V_{MC, TCID50/mL,0}$, $V_{MC, RNA/mL}(0) = V_{MC, RNA/mL,0}$ for the strains with wild-type HA and NA.
The initial values of all other compartments are 0.

For the single-cycle experiments, and for the multi-cycle experiments with PR8 HA and NA, the virus was incubated for an hour, then the cells were washed and the medium replaced.
For the strains with PR8 HA and NA, we model this by setting $T(-1) = T_0$, $V_{pfu/mL}(-1) = V_{pfu/mL,-1}$, and all other compartments to zero.
We then solve Eq.~\ref{eq:model_obs} from $t = -1$ to $t = 0$, set $V_{pfu/mL}(0) = V_{pfu/mL,0}$, then continue solving the equations until the end of the experiment.
For these experiments, all reported times are times post-washing.
The strains with wild-type HA and NA are modelled in the same way, but in units of TCID$_{50}$/mL, and with the addition of Eq.~\ref{eq:model_rna}.

For the mock-yield experiments, the initial conditions are $V_{mock, pfu/mL}(0) = V_{mock, pfu/mL,0}$ for the strains with PR8 HA and NA, and $V_{mock, TCID50/mL}(0) = V_{mock, TCID50/mL,0}$ for the strains with wild-type HA and NA (total virus was not measured for these experiments).
The initial values of all other compartments are 0.

Model equations are solved using the \texttt{odin} package version 0.1.1, available at \url{https://github.com/mrc-ide/odin}.

We model the observed viral load (infectious or total) as lognormally distributed around the true viral load.
For example, for a single data point for the total viral load in a multi-cycle experiment, we have
\begin{equation}
P(\hat{V}_{MC, RNA/mL}(t) |\boldsymbol{\theta}) = 
\frac{1}{\sqrt{2\sigma_{RNA}^2\pi} } \exp \left\{-\frac{\left[\log_{10}\hat{V}_{MC, RNA/mL}(t)-\log_{10}V_{MC, RNA/mL}(t,\boldsymbol{\theta})\right]^2}{2\sigma_{RNA}^2} \right\} 
\label{eq:likelihood2}
\end{equation}
where $\hat{V}_{MC, RNA/mL}(t)$ is the observed viral load at time $t$ and $V_{MC, RNA/mL}(t,\boldsymbol{\theta})$ is the true viral load at time $t$ according to the model parameters $\boldsymbol{\theta}$.

In addition, for the plaque assay we model an observation threshold of $\Theta = 10$ pfu/mL below which the viral load is treated as censored~\citep{Yan2019}.
We denote below-threshold measurements as 0, so $\hat{V}_{pfu/mL}(t)$ can take the value 0 or any value above (and including) $\Theta$. 
The likelihood of a single data point according to the plaque assay given model parameters is then given by Eq.~\ref{eq:likelihood}.

\begin{subequations}
\begin{align}
P(\hat{V}_{pfu/mL}(t) |\boldsymbol{\theta}) &= 
\begin{cases} 
\frac{1}{\sqrt{2\sigma_{pfu}^2\pi} } \exp \left\{-\frac{\left[\log_{10}\hat{V}_{pfu/mL}(t)-\log_{10}V_{pfu/mL}(t,\boldsymbol{\theta})\right]^2}{2\sigma_{pfu}^2} \right\}  & \text{if } \hat{V}_{pfu/mL}(t) \geq \Theta, \\
\bigints_0^\Theta\frac{1}{\sqrt{2\sigma^2\pi} } \exp \left\{-\frac{\left[\log_{10}x-\log_{10}V_{pfu/mL}(t,\boldsymbol{\theta})\right]^2}{2\sigma^2} \right\} dx & \text{if } \hat{V}_{pfu/mL}(t) = 0,\\
0 & \text{otherwise}.
\end{cases}
\end{align}
\label{eq:likelihood}
\end{subequations}
Errors are assumed to be independent, so the likelihood across all data points is the product of the likelihoods of each of the data points.

\section*{Model fitting}


The values of fixed model parameters are as follows.

\begin{tabular}{rrr}
  \hline
Parameter & Value & Units \\ 
  \hline
  $c_{RNA}$ & $10^{-3}$ & h$^{-1}$\\ 
  $n_L$ & 10 & -\\
  $n_I$ & 10 & -\\
  $T_0$ & $10^6$ (WT) or $2.5 \times 10^6$ (PR8) & cell\\
  $S$ & $10$ (WT) or $3$ (PR8) & mL\\
  $\Theta$ & 0 or 10 & pfu mL$^{-1}$\\
   \hline
\end{tabular}

The prior bounds for fitted parameters are as follows.  Parameters with Y in the log transform column have uniform priors in log space, while the others have uniform priors in linear space.  Priors are the same for the strains with wild-type HA and NA and the strains with PR8 HA and NA, except the units for infectious virus are pfu and TCID$_{50}$ respectively.
Also, the incubation period is absent for the multi-cycle experiments with wild-type HA and NA, so $V_{MC, TCID_{50}/mL, -1}$ is fixed at 0 for these strains.

\begin{tabular}{rrrr}
  \hline
Parameter & log transform & Prior bounds & Units \\ 
  \hline
  $c$ & Y & $[10^{-3}, 10^0]$ & h$^{-1}$\\ 
  $\beta_{inf}$ & Y & $[10^{-10}, 10^{-1}]$ & infectious unit $^{-1}$ h$^{-1}$ \\
  $\tau_L$ & N & $[0.1, 12]$ & h \\ 
  $\tau_I$ & Y & $[10^0, 10^3]$ & h \\ 
  $p_{SC, pfu/mL}/p_{MC, pfu/mL}$ & Y & $[10^{-5}, 10^0]$ & -\\ 
  $p_{MC, pfu/mL}$ & Y & $[10^{-2}, 10^3]$ & pfu mL$^{-1}$ cell$^{-1}$ h$^{-1}$ \\ 
  $p_{RNA/mL}$ & Y & $[10^0, 10^6]$ & copy number mL$^{-1}$ cell$^{-1}$ h$^{-1}$ \\ 
  $n$ & Y & $[10^0, 10^7]$ & infectious unit pfu$^{-1}$\\ 
  $V_{SC, pfu/mL, -1}$ & Y & $[10^0, \frac{10^{10}T_0}{S}]$ & pfu mL$^{-1}$\\
  $V_{SC, pfu/mL, 0}$ & Y & $[10^{-2},10^7]$ & pfu mL$^{-1}$\\
  $V_{SC, RNA/mL, 0}/V_{SC, pfu/mL, 0}$ & Y & $[10^0,10^6]$ & copy number pfu$^{-1}$\\
  $V_{MC, pfu/mL, -1}$ & Y & $[10^0,\frac{T_0}{10S}]$ & pfu mL$^{-1}$\\
  $V_{MC, pfu/mL, 0}$ & Y & $[10^{-2}, 10^4]$ & pfu mL$^{-1}$\\
  $V_{MC, RNA/mL, 0}/V_{MC, pfu/mL, 0}$ & Y & $[10^0, 10^6]$ & copy number pfu$^{-1}$\\
  $V_{mock, pfu/mL, 0}$ & Y & $[10^5, 10^{10}]$ & pfu mL$^{-1}$\\
  $\sigma_{pfu}$ & N & $[0, 2]$ & pfu mL$^{-1}$\\
  $\sigma_{RNA}$ & N & $[0, 2]$ & copy number mL$^{-1}$\\
   \hline
\end{tabular}

The joint posterior distribution of model parameters is obtained using an adaptive Metropolis-Hastings algorithm, which we implemented in \R as the package \texttt{lazymcmc}, available at \url{https://github.com/ada-w-yan/lazymcmc/}.
Parallel tempering (as developed by Geyer \textit{et al.}~\citep{Geyer1991} and reviewed by Earl2005 \textit{et al.}~\citep{Earl2005}) was implemented to improve exploration of parameter space.
Five parallel chains with different temperatures were used; the temperatures were calibrated as previously described in Yan \textit{et al.}~\citep{Yan2019}.
To assess convergence, three such sets of five parallel chains were run, after which convergence was assessed using the \texttt{coda} package version 0.19-1~\citep{coda}, as previously described in Yan \textit{et al.}~\citep{Yan2019}.

Code to reproduce all results can be found at \repo{}.

\section*{Calculation of the proportion of infectious virions in each generation}

To calculate the proportion of infectious virions in each generation, we modify Eq.~\ref{eq:model_obs} to track the generation number of latent cells, infectious cells and infectious virions.

\begin{subequations}
\begin{align}
\frac{dT}{dt} &= -\beta T \sum_{g = 1}^G V_{pfu/mL, g},&\\
\frac{dL_{1, g}}{dt} &= \beta T V_{pfu/mL, g} - \frac{n_{L}}{\tau_{L}} L_{1, g}, &g = 1, \dots, G,\\
\frac{dL_{i, g}}{dt} &= \frac{n_{L}}{\tau_{L}} (L_{i-1, g} - L_{i, g}), &g = 1, \dots, G,\\
\frac{dI_{1, g}}{dt} &= \frac{n_{L}}{\tau_{L}} L_{i, g} - \frac{n_{I}}{\tau_{I}} I_{1, g}, & = 1, \dots, G,\\
\frac{dI_{i, g}}{dt} &= \frac{n_{I}}{\tau_{I}} (I_{i-1, g} - I_{i, g}), &g = 1, \dots, G,\\
\frac{dV_{pfu/mL}, 1}{dt} &= - c V_{pfu/mL, 1} - \beta_{inf} T V_{pfu/mL, 1},&\\
\frac{dV_{pfu/mL}, g}{dt} &= p \sum_{i=1}^{n_{I}} I_{i, g - 1} - c V_{pfu/mL, g} &\nonumber\\
&- \beta_{inf} T V_{pfu/mL, g}, &g = 2, \dots, G-1,\\
\frac{dV_{pfu/mL}, G}{dt} &= p \sum_{i=1}^{n_{I}} (I_{i, G - 1} + I_{i, G}) - c V_{pfu/mL, G}&\nonumber\\
& - \beta_{inf} T V_{pfu/mL, G}.&
\end{align}
\label{eq:model_gen}
\end{subequations}
Here, $g$ denotes the generation number, and $G$ is the maximum number of generations tracked, which is capped for computational purposes at $G = 20$.
The inoculum is defined as the first generation of virions, which infect cells to produce first-generation latent cells, which become first-generation infectious cells, which produce second-generation virions, and so forth.
We solve the equations for the maximum likelihood parameters for a given strain, using multi-cycle initial conditions, to determine the peak viral load and infectious virion distribution at that time.

\section*{Varying cellular infection parameters systematically for the generation number model}

To create Fig.~4, first, the cellular generation time ($T_G$) was held constant while the cellular reproduction number ($R_0$) and the initial growth rate ($r$) were increased by increasing $p_{inf,MC}$, the production rate of infectious virions for a multi-cycle experiment.
Second, $R_0$ was held constant while $T_G$ was increased and $r$ was decreased.  This was accomplished by increasing the infectious cell lifespan $\tau_I$ and decreasing $p_{inf,MC}$ simultaneously to hold the burst size $p_{inf,MC}\tau_I$ constant.
Third, $r$ was held constant while $R_0$ and $T_G$ were increased.  This was accomplished by increasing the infectious cell lifespan $\tau_I$ and calculating the value of $p_{inf,MC}$ to keep $r$ constant.
The value of $p_{inf,MC}$ with the desired value of $r$ was found using the \texttt{uniroot} function in \R{}.

\section*{Fixed generation distribution}

To create Fig.~S8, we simulated from the model by Russell \textit{et al.}~\citep{Russell2012}, where one starts with an initial number of virions, and at fixed intervals (the generation time), each virion produces $R_0$ secondary virions and the original virions are discarded, advancing the model by one generation.
In this model, the viral load is capped at $V_{max} = 10^{14}$ virions.
Starting from one initial virion, we simulate viral growth until the viral load reaches $V_{max}$, at which we declare the peak is reached and record the number of generations taken to get there.
For this model, the relationship between the cellular reproduction number $R_0$, the cellular generation time $T_G$ and the initial growth rate $r$ is

\begin{equation}
R_0 = \exp(rT_G).
\label{eq:R_0_T_G_r_fixed}
\end{equation}

First, we held $T_G$ constant at 12 h while varying $R_0$ between $10^3$ and $10^6$, calculating the corresponding value of $r$ using Eq.~\ref{eq:R_0_T_G_r_fixed}.
Then, we held $R_0$ constant at $10^4$ while varying $T_G$ between 6 and 24 h, calculating the corresponding value of $r$ using Eq.~\ref{eq:R_0_T_G_r_fixed}.
Last, we held $r$ constant at 1 while varying $T_G$ between 6 and 24 h, calculating the corresponding value of $R_0$ using Eq.~\ref{eq:R_0_T_G_r_fixed}.

\section*{Model ignoring the loss of virus due to entry into target cells}

To ignore the loss of virus due to entry into target cells, we modify Eq.~\ref{eq:model_obs} by setting $\beta_{inf}$ to 0.
We decouple the relationship between $\beta_{inf}$ and $\beta_{pfu/mL}$, and fit $\beta_{pfu/mL}$ directly.
The prior bounds for $\beta_{pfu/mL}$ are $[10^{-10}, 10^{-1}]$.

\section*{Model where the production rate of virions increases over an infectious cell's age of infection}

We modify Eq.~\ref{eq:model_obs} such that the production rate of virions depends on the age of the infectious cell:
\begin{subequations}
\begin{align}
\frac{dT}{dt} &= -\beta_{pfu/mL} T V_{pfu/mL},\\
\frac{dL_{1}}{dt} &= \beta_{pfu/mL} T V_{pfu/mL} - \frac{n_{L}}{\tau_{L}} L_{1},\\
\frac{dL_{i}}{dt} &= \frac{n_{L}}{\tau_{L}} (L_{i-1} - L_{i}),\\
\frac{dI_{1}}{dt} &= \frac{n_{L}}{\tau_{L}} L_{i} - \frac{n_{I}}{\tau_{I}} I_{1},\\
\frac{dI_{i}}{dt} &= \frac{n_{I}}{\tau_{I}} (I_{i-1} - I_{i}),\\
\frac{dV_{pfu/mL}}{dt} &= \sum_{i=1}^{n_{I}}\frac{i}{n_I}p_{pfu/mL,max} I_{i} - c V_{pfu/mL} - \beta_{inf} T V_{pfu/mL},\\
\frac{dV_{RNA/mL}}{dt} &= \sum_{i=1}^{n_{I}} \frac{i}{n_I}p_{RNA/mL,max}I_{i} - c_{RNA} V_{RNA/mL} - \beta_{inf}n T V_{pfu/mL}.
\end{align}
\label{eq:model_obs_linear}
\end{subequations}

We fit $p_{MC,pfu/mL,max}$ and $p_{RNA/mL,max}$ using the prior bounds $[10^{-2}, 10^{10}]$ and $[10^0, 10^{10}]$ respectively.
As per the model in the main text, we assume that the maximum infectious virion production rate for the single-cycle experiment is lower or equal to that of the multi-cycle experiment, and fit $p_{SC,pfu/mL,max}/p_{MC,pfu/mL,max}$ with prior bounds $[10^{-5}, 10^0]$.

\clearpage

\section*{Supporting Figures}

\renewcommand\thefigure{S\arabic{figure}}   
\setcounter{figure}{0}

\begin{figure}[h]
\centering
\includegraphics[width = \textwidth]{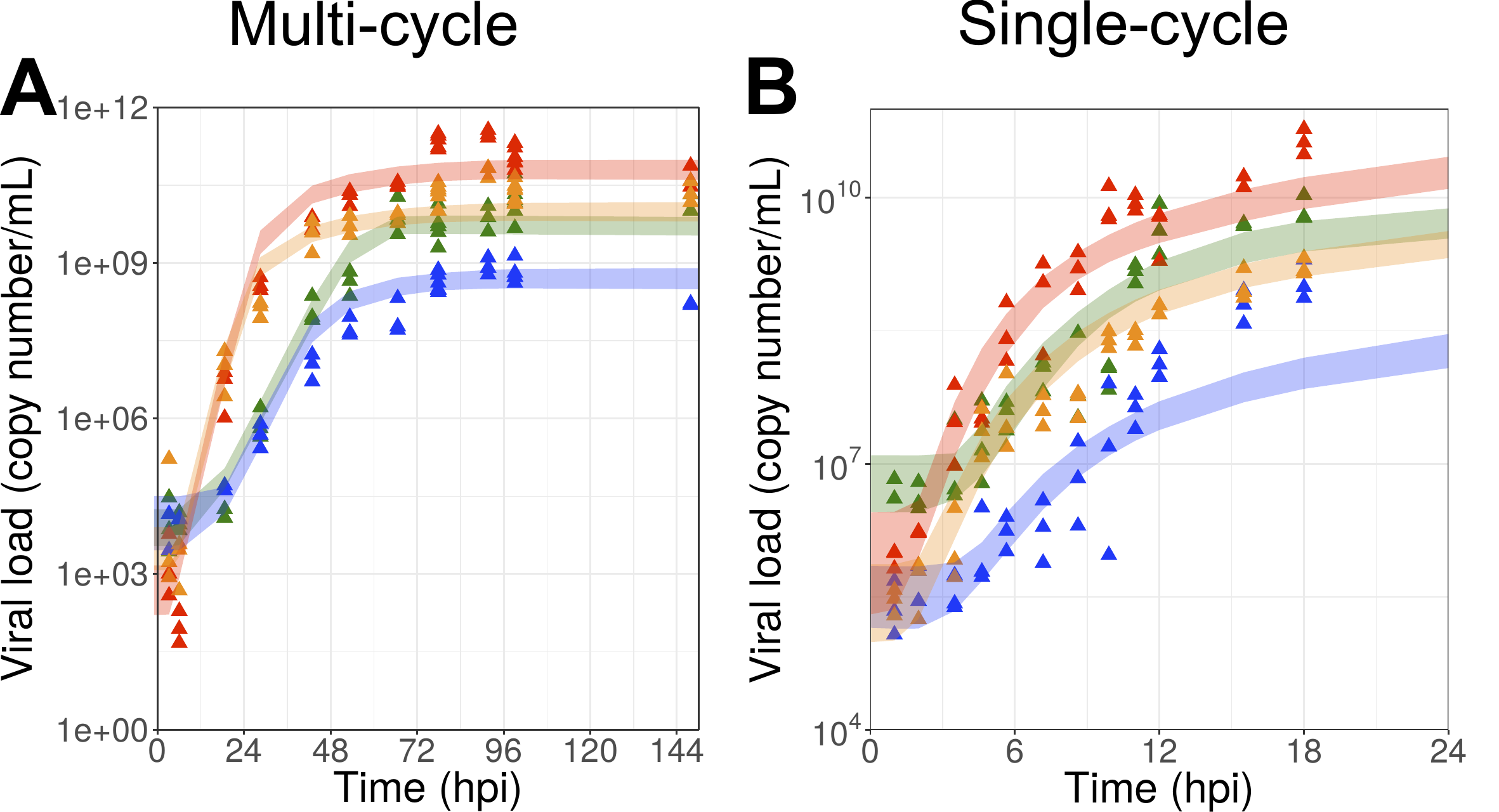}
\caption{The total viral load for the strains with wild-type HA and NA.
Fitted 95\% credible intervals are shown as shaded areas on top the the data (triangles).}
\label{fig:total_viral_load}
\end{figure}

\begin{figure}[t]
\includegraphics[width = \textwidth]{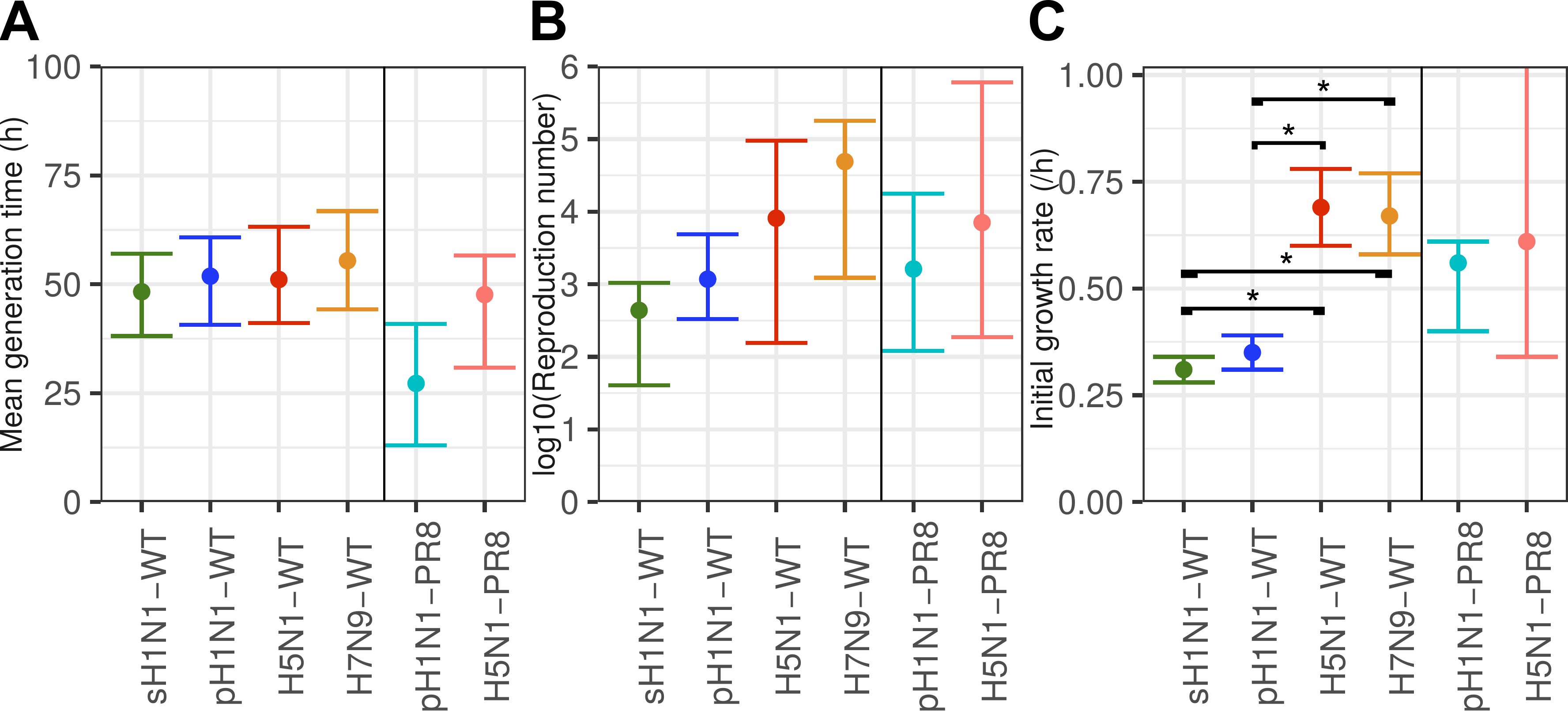}
\caption{
\textbf{Parameter estimates using the multi-cycle data only.}
The median and 95\% credible intervals for (A) the mean generation time, (B) cellular reproduction number, and (C) initial growth rate, for the WT strains (left of each panel), and the strains with PR8 HA and NA (right).
Asterisks denote $p < 0.001$.
}
\label{fig:summary_stats_mc}
\end{figure}

\begin{figure}[h]
\centering
\includegraphics[width = \textwidth]{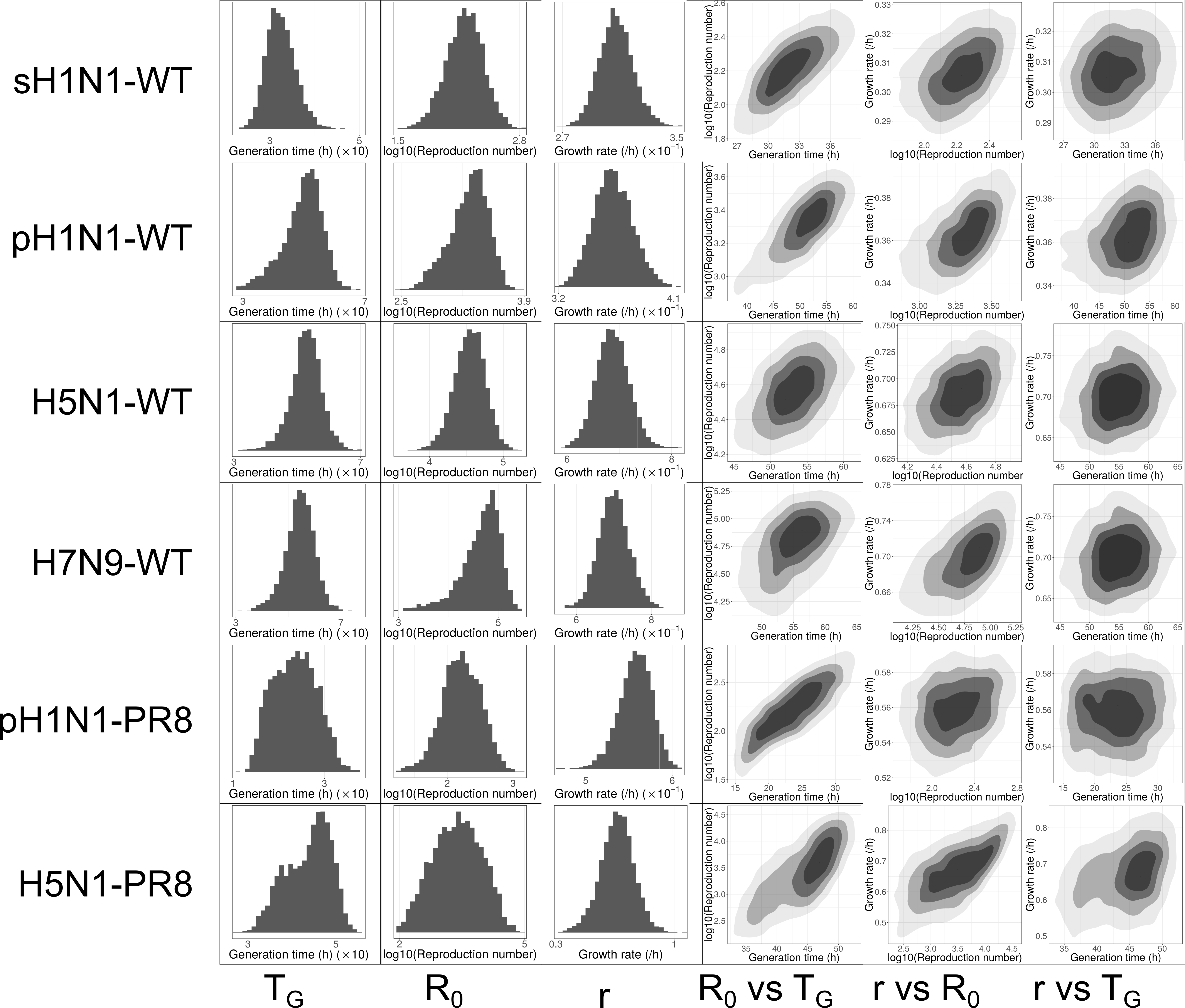}
\caption{Left three columns: Posterior distributions for (left to right) the mean generation time $T_G$, the basic reproduction number $R_0$, and the initial growth rate $r$.  Right columns: joint posterior density of $R_0$ versus $T_G$, $r$ versus $R_0$, and $r$ versus $T_G$.
}
\label{fig:bivariate_sum_stats}
\end{figure}

\begin{figure}[t]
\includegraphics[width = \textwidth]{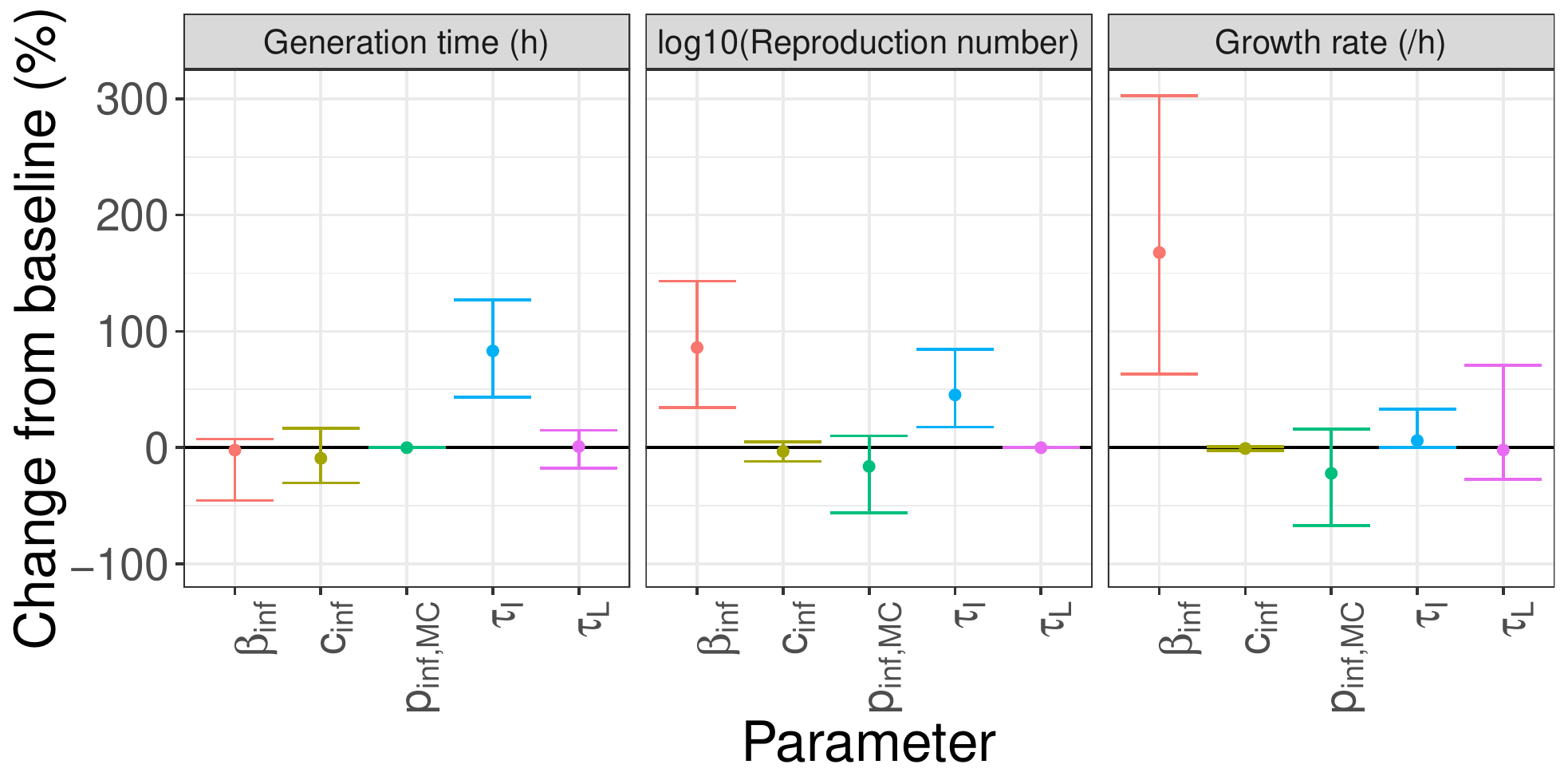}
\caption{
\textbf{Percentage change in the cellular infection parameters --- basic reproduction number, mean generation time and initial growth rate --- for sH1N1-WT, changing the value of the parameter in the horizontal axis to that of H7N9-WT.}
The baseline parameters are sampled from the joint posterior distribution for sH1N1-WT.
As a sensitivity analysis, we change the rates of infection processes in the model one by one to rates sampled from the joint posterior distribution for H7N9-WT, to determine each rate's influence on the value of the cellular infection parameters.
These rates are changed by changing $\beta_{inf}$, the infectivity of virions; $c_{inf}$, the rate at which virions lose infectivity; $p_{inf,MC}$, the production rate of infectious virions in the multi-cycle experiment; $\tau_I$, the mean infectious period; and $\tau_L$, the mean latent period (as defined in the model equations).
For example, the leftmost bar shows the percentage change in the mean generation time for sH1N1-WT, if the infectivity of virions $\beta_{inf}$ were changed to that of H7N9-WT.  The value of this bar is determined by calculating the mean generation time for sH1N1-WT; taking the rates of infection processes for sH1N1-WT, and replacing the infectivity $\beta_{inf}$ with the value for H7N9-WT; re-calculating the mean generation time; and finding the percentage change between the two mean generation time values.  
This process is repeated for many pairs of samples from the joint posterior distributions for sH1N1-WT and H7N9-WT, to obtain the median and 95\% credible intervals plotted here.
}
\label{fig:change_single_par}
\end{figure}

\begin{figure}[t]
\includegraphics[width = \textwidth]{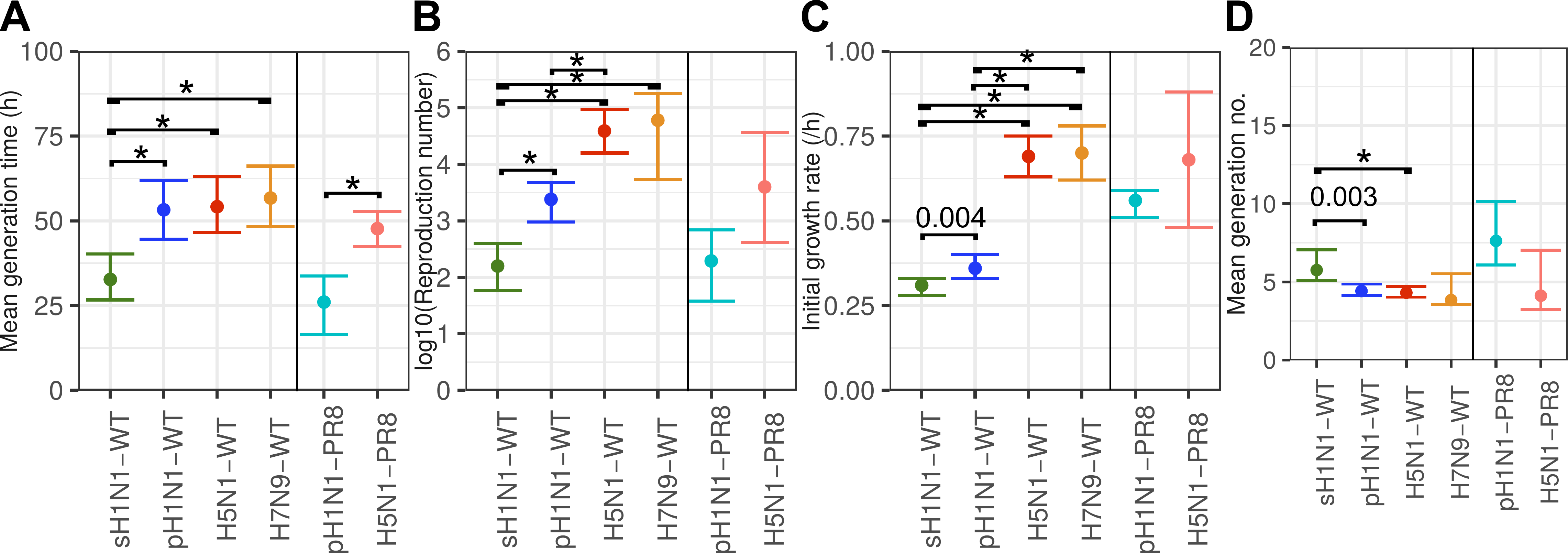}
\caption{
\textbf{Parameter estimates and correlations for a model ignoring loss of free virions due to entry into target cells.}
The median and 95\% credible intervals for (A) the mean generation time, (B) basic reproduction number, (C) initial growth rate and (D) mean generation number, for the WT strains (left of each panel), and the strains with PR8 HA and NA (right).
Statistically significant pairs are labelled ($\alpha = 0.05$ with Bonferroni correction for seven pairwise tests per parameter).
Asterisks denote $p < 0.001$.
}
\label{fig:loss}
\end{figure}

\begin{figure}[t]
\includegraphics[width = \textwidth]{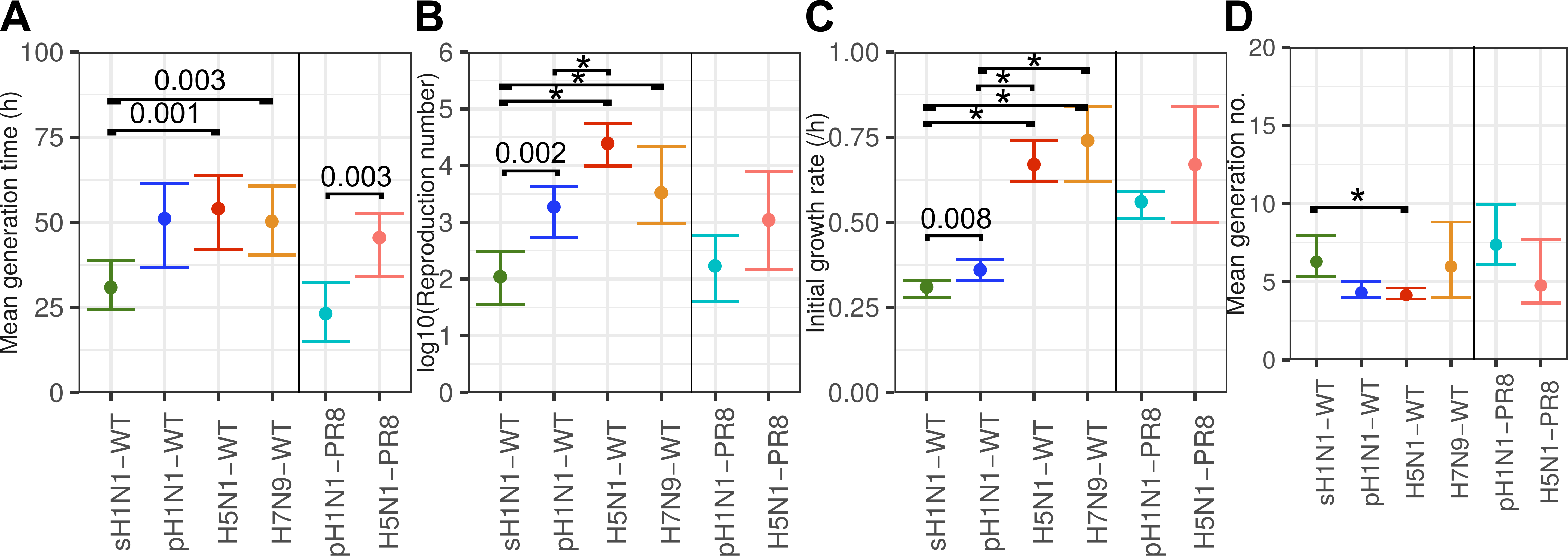}
\caption{
\textbf{Parameter estimates and correlations for a narrower distribution of the latent period and duration of virion production, setting $n_L = n_I = 60$ in the model equations.}
The median and 95\% credible intervals for (A) the mean generation time, (B) basic reproduction number, (C) initial growth rate and (D) mean generation number, for the WT strains (left of each panel), and the strains with PR8 HA and NA (right).
Statistically significant pairs are labelled ($\alpha = 0.05$ with Bonferroni correction for seven pairwise tests per parameter).
Asterisks denote $p < 0.001$.
}
\label{fig:60}
\end{figure}

\begin{figure}[h]
\centering
\includegraphics[width = .8\textwidth]{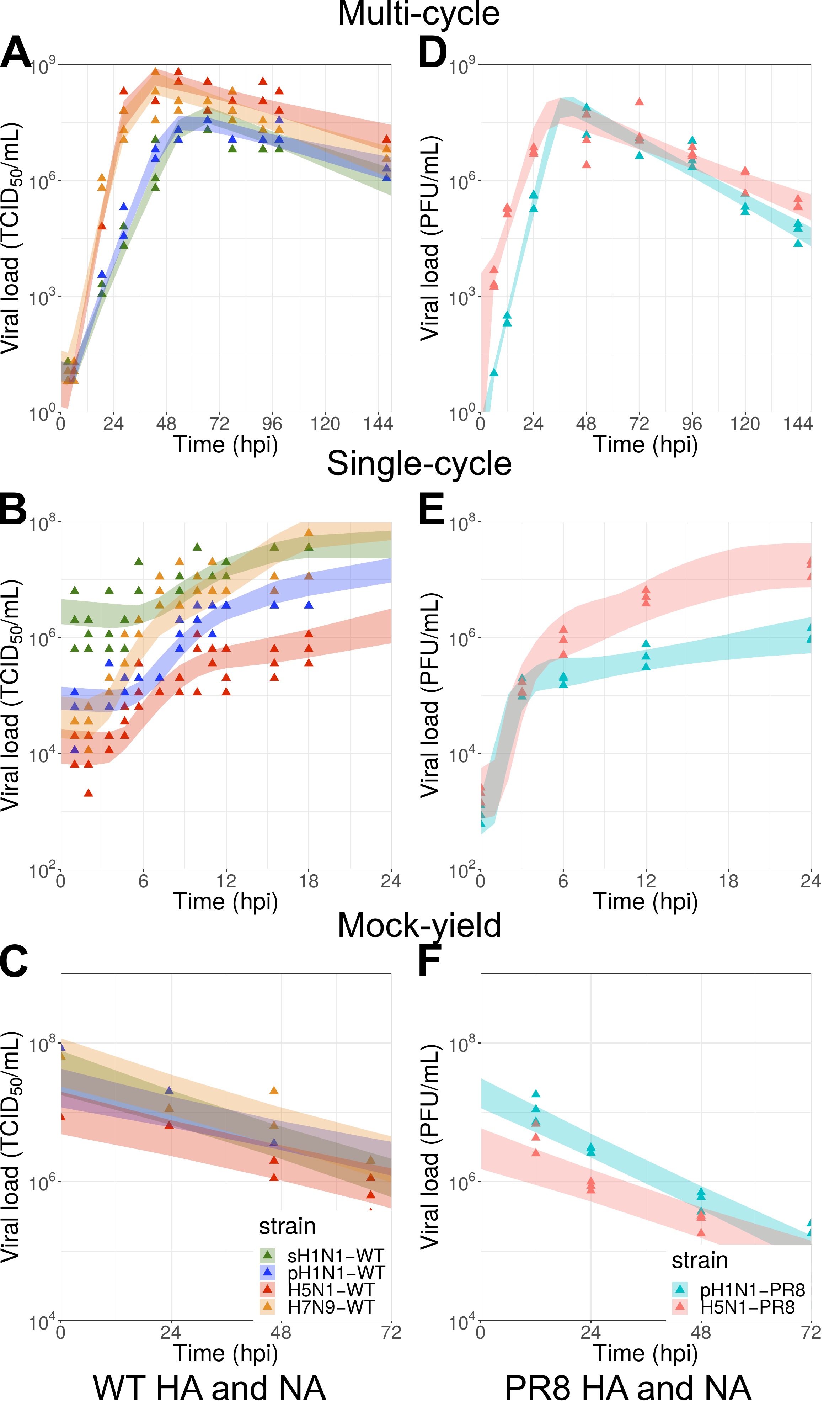}
\caption{
\textbf{Model with linearly increasing production rate: fits to experimental data.}
The infectious viral load for the WT strains (A-C) and the strains with PR8 HA and NA (D-F). Fitted 95\% credible intervals are shown as shaded areas on top the the data (triangles). The infectious viral load is shown for (from top) the multi-cycle experiments, the single-cycle experiments and the mock-yield experiments.}
\label{fig:data_linear}
\end{figure}

\begin{figure}[t]
\centering
\includegraphics[width = \textwidth]{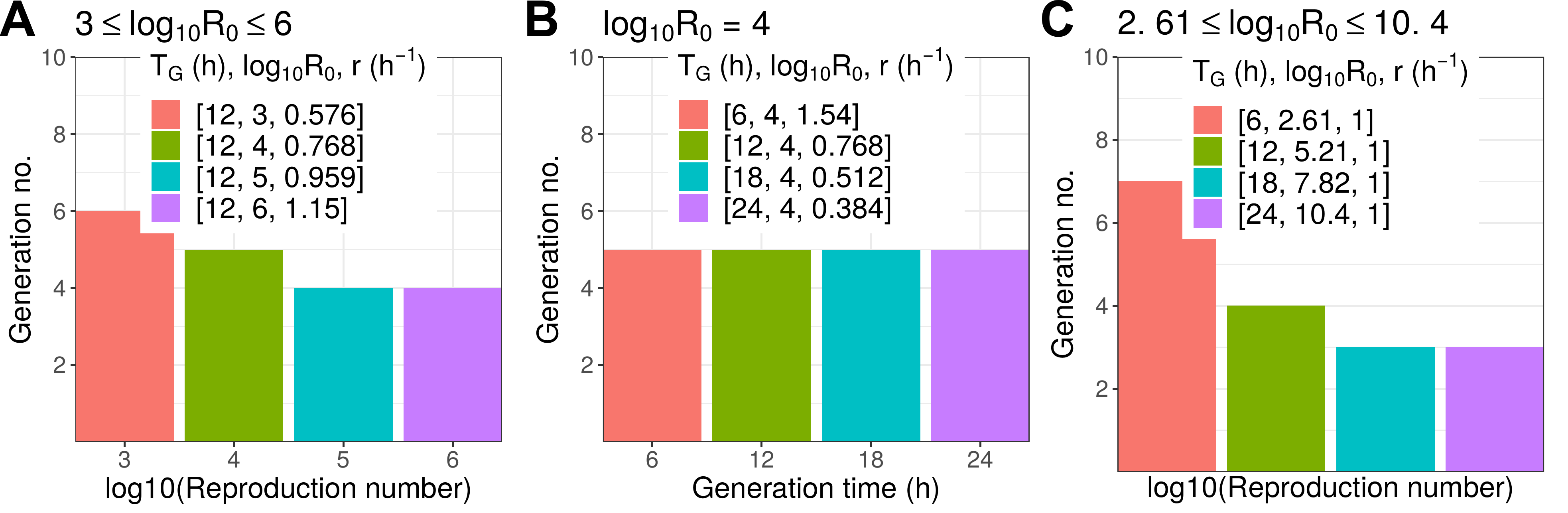}%
\caption{
\textbf{Simulated generation number of virions, as cellular infection parameters are changed systematically, for a model with synchronised generations.} The generation to which the virions belonged was calculated at the time of peak infectious viral load.
The cellular infection parameters in the legend were changed to the values shown.  In each panel, one parameter is held constant: the mean generation time $T_G$ (A), basic reproduction number $R_0$ (B), or initial growth rate $r$ (C).
}
\label{fig:gen_distribution_fixed}
\end{figure}

\clearpage
\bibliographystyle{apalike}
\bibliography{Mendeley,unpublished}